%% file: ms.tex
\documentclass[preprint,12pt,authoryear]{elsarticle}

\usepackage[utf8]{inputenc}
\DeclareUnicodeCharacter{223C}{~}
\DeclareUnicodeCharacter{00B0}{}
\usepackage{amsmath,amssymb}
\usepackage{cases}
\usepackage{graphicx}
\usepackage{stackrel}
\usepackage{float}
\usepackage[version=4]{mhchem}
\usepackage[font=small, labelfont=bf, justification=justified, singlelinecheck=false]{caption}
\usepackage{subcaption}
\usepackage[overload]{empheq}

\usepackage{subfiles}

\newcommand{\fth}{$f_{\mathrm{T}}(\mathrm{H}_2)$}
\newcommand{\ftvh}{$f^{\mathrm{v}}_{\mathrm{T}}(\mathrm{H}_2)$}
\newcommand{\fo}[1]{$f\mathrm{O}_2$}
\newcommand{\hflux}{$\Phi_{\mathrm{volc}}(\ce{H2})$}
\newcommand{\percmpers}{$\mathrm{cm}^{-1}\,\mathrm{s}^{-1}$}
\newcommand{\percmsqpers}{$\mathrm{cm}^{-2}\,\mathrm{s}^{-1}$}

\journal{Earth and Planetary Science Letters}

\begin{document}

\begin{frontmatter}



\title{Can Volcanism Build Hydrogen-Rich Early Atmospheres?}


\author[label1]{Philippa Liggins}
\ead{pkl28@cam.ac.uk}
\author[label1,label2]{Oliver Shorttle}
\author[label1,label3,label4]{Paul B. Rimmer}

\address[label1]{Department of Earth Sciences, University of Cambridge, Cambridge, CB2 3EQ, UK}
\address[label2]{Institute of Astronomy, University of Cambridge, Cambridge, CB3 0HA, UK}
\address[label3]{Cavendish Astrophysics, University of Cambridge, JJ Thomson Avenue, Cambridge CB3 0HE, UK}
\address[label4]{MRC Laboratory of Molecular Biology, Francis Crick Ave, Cambridge CB2 0QH, UK}

\begin{abstract}
Hydrogen in rocky planet atmospheres has been invoked in arguments for extending the habitable zone via \ce{N2}-\ce{H2} and \ce{CO2}-\ce{H2} greenhouse warming, and providing atmospheric conditions suitable for efficient production of prebiotic molecules.
On Earth and Super-Earth-sized bodies, where hydrogen-rich primordial envelopes are quickly lost to space, volcanic outgassing can act as a hydrogen source, provided it balances the hydrogen loss rate from the top of the atmosphere. 
Here, we show that both Earth-like and Mars-like planets can sustain atmospheric \ce{H2} fractions of several percent across relevant magmatic \fo{} ranges. In general this requires hydrogen escape to operate somewhat less efficiently than the diffusion limit.
We use a thermodynamical model of magma degassing to determine which combinations of magma oxidation, volcanic flux and hydrogen escape efficiency can build up appreciable levels of hydrogen in a planet's secondary atmosphere.
On a planet similar to the Archean Earth and with a similar magmatic \fo{}, we suggest that the mixing ratio of atmospheric \ce{H2} could have been in the range 0.2-3\%, from a parameter sweep over a variety of plausible surface pressures, volcanic fluxes, and \ce{H2} escape rates.
A planet erupting magmas around the Iron-W{\"u}stite (IW) buffer (i.e., $\sim$3 log \fo{} units lower than the inferred Archean mantle \fo{}), but with otherwise similar volcanic fluxes and \ce{H2} loss rates to early Earth, could sustain an atmosphere with approximately 10-20\% \ce{H2}.
For an early Mars-like planet with magmas around IW, but a lower range of surface pressures and volcanic fluxes compared to Earth, an atmospheric \ce{H2} mixing ratio of $\sim$2-8\% is possible. On early Mars, this \ce{H2} mixing ratio could be sufficient to deglaciate the planet. However, the sensitivity of these results to primary magmatic water contents and volcanic fluxes show the need for improved constraints on the crustal recycling efficiency and mantle water contents of early Mars.
\end{abstract}


\end{frontmatter}


\section{Introduction}
Atmospheric \ce{H2} could be key to the development of terrestrial planets.
The presence of significant hydrogen fractions has been invoked in extending the habitable zone \citep[e.g.,][]{sagan1977ReducingGreenhousesTemperature,stevenson1999LifesustainingPlanetsInterstellar,wordsworth2012TransientConditionsBiogenesis, pierrehumbert2011HydrogenGreenhousePlanets,abbot2015PROPOSALCLIMATESTABILITY,ramirez2017VolcanicHydrogenHabitable} and providing conditions suitable for the production of prebiotic molecules via \ce{H2} buildup then loss \citep{miller1959OrganicCompoundSynthesis}.
Several studies have looked at the climatic consequences of adding a hydrogen fraction to the atmospheres of the early Earth \citep{wordsworth2013HydrogenNitrogenGreenhouseWarming}, early Mars \citep{ramirez2014WarmingEarlyMars,batalha2015TestingEarlyMars,wordsworth2017TransientReducingGreenhouse, hayworth2020WarmingEarlyMarsa} and terrestrial planets more widely \citep[e.g.,][]{ramirez2017VolcanicHydrogenHabitable}.
Others have looked at how hydrogen can be introduced into the earliest primary atmospheres of terrestrial bodies from proto-planetary nebular gases \citep{stevenson1999LifesustainingPlanetsInterstellar}, magma oceans \citep{deng2020MagmaOceanOrigin} and accretion/impact processes \citep{schaefer2007OutgassingOrdinaryChondritic, schaefer2010ChemistryAtmospheresFormed}.
However, hydrogen is a light element which escapes easily to space; on terrestrial sized bodies, any primary atmosphere should be blown off by high rates of stellar irradiation within the first few million years \citep{pierrehumbert2011HydrogenGreenhousePlanets}.
Therefore, to maintain a hydrogen-rich atmosphere on a rocky planet, \ce{H2} must be constantly replenished over geologic timescales to offset its continuous loss to space.

In the presence of life, methanogens and anoxygenic phototrophs can be an important source of hydrogen in the form of methane \citep{kharecha2005CoupledAtmosphereEcosystem}.
However, in a prebiotic context geologic sources of \ce{H2} are required to generate abiotic secondary atmospheres (i.e., atmospheres generated post-accretion and magma ocean phase, and before the emergence of life).
These can include the serpentinisation of ultramafic rock, metamorphic fluxes of \ce{CH4} from hydrothermal vents, and volcanism \citep{kasting2013WhatCausedRise}. It is this latter possibility of volcanic \ce{H2} production in an early secondary atmosphere that this paper focuses on.

A key parameter determining the build up of \ce{H2} in planetary atmospheres is the escape rate of hydrogen to space.
Were this loss rate to have been lower on the early Earth, moderate to significant fractions of hydrogen could have built up \citep{tian2005HydrogenRichEarlyEarth,kuramoto2013EffectiveHydrodynamicHydrogen}.
In these studies, the hydrogen source is assumed to be volcanic and is based on extrapolating from the modern volcanic hydrogen flux.
The missing link for these studies is a more thorough analysis of the amount of hydrogen which could reasonably be released to an atmosphere through volcanism, as a function of planetary interior and exterior conditions.
There are 3 primary controls on this; 1) the redox state of the degassing magma, 2) the pressure of degassing, and 3) the overall volcanic flux of the planet.
Once these factors have been constrained, the hydrogen loss rate necessary to achieve a particular hydrogen fraction from volcanism can be evaluated.
Here, we aim to address this issue by using a magma degassing code to constrain the possible range of volcanically sustained \ce{H2} in planetary atmospheres, linking the atmospheric \ce{H2} fraction to the potential oxidation state, geological activity and atmospheric loss rates of early Earth, early Mars and terrestrial planets.

\section{Controls on a volcanic hydrogen atmospheres}

\subsection{Magma degassing}
Volcanic gases are widely considered to be approximately in equilibrium with the oxidation state of the magma from which they exsolved \citep[e.g.,][]{moretti2004OxidationStateVolatile,burgisser2015SimulatingBehaviorVolatiles}.
Therefore, to change the speciation of a gas phase from oxidised (\ce{H2O}, \ce{CO2} and \ce{SO2} dominated) to reduced (\ce{CO}, \ce{H2} and \ce{H2S} dominated), and therefore more \ce{H2}-rich, the \fo{} of the magma needs to be lower. 
A greater dissolved water content at a constant \fo{} will result in a melt with a greater total atomic hydrogen content, and therefore will also increase the volcanic hydrogen emission.

Pressure exerts a dominant control over volatile partitioning between the melt and the gas phase in a magmatic system, and therefore also controls the composition of a volcanic gas phase \citep{gaillard2014TheoreticalFrameworkVolcanic}.
Species with a low solubility such as \ce{CO2} and \ce{CO} exsolve at higher pressures and therefore dominate the early stages of magma degassing, with more soluble and often more abundant phases such as water becoming dominant at low pressures \citep[e.g.,][]{dixon1995ExperimentalStudyWater}.
The effect of pressure on the composition of secondary volcanic atmospheres has been examined by \cite{gaillard2014TheoreticalFrameworkVolcanic}, in which they assume average redox conditions across the solar system driven by graphite saturation buffering an ascending magma’s \fo{},

\begin{equation}
\underset{\mathrm{graphite}}{\ce{C}} +  \underset{\mathrm{gas}}{\ce{O2}} \leftrightharpoons \underset{\mathrm{gas}}{\ce{CO2}},
\end{equation}
and therefore perform all their calculations starting at FMQ\,-1.5.
However, the average \fo{} of erupting MORB on Earth (a relatively oxidised planet) is FMQ\,-0.41$\pm$0.43 \citep{bezos2005Fe3SFeRatios}, and values across the solar system vary down to IW-5 on Mercury \citep{zolotov2013RedoxStateFeO, wadhwa2008RedoxConditionsSmall}.
Given the similar potential for \fo{} diversity among exoplanets \citep{doyle2019OxygenFugacitiesExtrasolar}, here we examine a wide range of \fo{} values.

While pressure and magmatic \fo{} can control the mass and composition of the gas phase, the total amount of \ce{H2} emitted to an atmosphere will also be controlled by the volcanic flux.
A higher flux can be achieved either through a larger total volatile content of the magma, which then releases a greater volume of gas into the atmosphere per unit mass of magma degassed, or a greater total flux of magma to the surface, such as may occur in a younger volcanically active planet, or a planet experiencing tidal heating (e.g., Io in our own solar system).

\subsection{Hydrogen escape mechanisms}

The escape flux of hydrogen from an atmosphere is dictated by the rate at which hydrogen diffuses through the layer of background air between the homopause (the level below which an atmosphere is well mixed) and the exobase, above which the atmosphere becomes collisionless.
I.e., the escape flux is limited by the rate at which \ce{H2} can be supplied to the altitude where it can escape Earth's atmosphere.
In a diffusion-limited escape regime, once hydrogen diffuses above the exobase it is rapidly lost from the atmosphere so that the rate of replenishment from below is the limiting factor.
These removal processes can be thermal, with two end-member approximations, or involve suprathermal mechanisms (e.g., photochemical escape, charge exchange and polar wind) which we do not consider here.

Thermal escape can occur as Jeans' escape; where molecules from the high-energy tail of the molecular thermal distribution attain enough kinetic energy to escape from the exobase; and hydrodynamic escape, occurring when heating of the atmosphere below the exobase by stellar extreme ultraviolet radiation (XUV) causes an upward pressure gradient driving a bulk, radial outflow.
Jeans' escape occurs with a high temperature exobase, and is responsible for a non-negligible fraction of hydrogen escape occurring today on Earth and Mars \citep{catling2017AtmosphericEvolutionInhabited}.
Hydrodynamic escape becomes dominant when the XUV flux is sufficient to drive a bulk outflow, and is key early in the lifetime of planetary systems, when the parent star is more active \citep{tu2015ExtremeUltravioletXray}.

Hydrogen escape from Earth is generally assumed to be diffusion-limited, and this diffusion rate can be linked to the concentration of hydrogen in an atmosphere \citep{walker1977EvolutionAtmosphere}.
However, some models of the early atmosphere suggest a cold exobase, and insufficient stellar XUV, resulting in a less efficient, energy limited escape regime \citep{tian2005HydrogenRichEarlyEarth, kuramoto2013EffectiveHydrodynamicHydrogen}.
Estimates for the hydrogen mixing ratio in the Archean atmosphere range from $<$0.1\,\% \citep[e.g.,][]{walker1977EvolutionAtmosphere,kasting1993EarthEarlyAtmosphere}, through $\leq$1\,\% \ce{H2} \citep{kuramoto2013EffectiveHydrodynamicHydrogen,zahnle2019StrangeMessengerNew} up to $~$30\,\% \citep{tian2005HydrogenRichEarlyEarth}.
To reflect uncertainty around the escape rate of \ce{H2}, we model cases both at and below the diffusion limit in this paper.
This is initially carried out in manner which is agnostic of the specific physics of hydrogen escape, so that the reduction from the diffusion limit into an energy limited scenario can be applied to planets generally, including those in different stellar regimes (and therefore different high energy UV spectra). However, escape fluxes below the diffusion limited rate are later compared to XUV-limited escape to physically inform results for the early Earth and Mars.

\section{Modelling Volcanic Outgassing} \label{volc_outgassing}
\subsection{Thermodynamics}
To calculate the gases input to an atmosphere from a volcanic source, we built a model using the mass balance and equilibrium constants method \citep[e.g.,][]{holloway1987IgneousFluidsThermodynamic, gaillard2014TheoreticalFrameworkVolcanic, burgisser2015SimulatingBehaviorVolatiles}. This model calculates the speciation and volume of a COHS gas phase in equilibrium with a silicate melt at a given pressure, temperature and magma \fo{}, considering both the homogeneous gas-phase equilibria and the heterogeneous gas-melt equilibria.
Equilibrium is always assumed throughout the system, and we run the calculations at the prescribed surface pressure and eruptive \fo{}.
The COHS volatile system is made up of 9 species, undergoing the following reactions,

\begin{equation}
	\ce{H2} + \frac{1}{2}\ce{O2} \leftrightharpoons \ce{H2O}
	\label{eqn:k1}
\end{equation}

\begin{equation}
	\ce{CO} + \frac{1}{2}\ce{O2} \leftrightharpoons \ce{CO2}
	\label{eqn:k2}
\end{equation}

\begin{equation}
	\ce{CH4} + 2\ce{O2} \leftrightharpoons \ce{CO2} + 2\ce{H2O}
	\label{eqn:k3}
\end{equation}

\begin{equation}
	\ce{H2S} + \frac{1}{2}\ce{O2} \leftrightharpoons \frac{1}{2}\ce{S2} + \ce{H2O}
	\label{eqn:k4}
\end{equation}

\begin{equation}
	\frac{1}{2}\ce{S2} + \ce{O2} \leftrightharpoons \ce{SO2}.
	\label{eqn:k5}
\end{equation}
Melt-gas equilibria are considered in the form of a solubility law, where the amount of species $x$ dissolved in the melt is found by applying a power law

\begin{equation}
    x^{\mathrm{melt}} = a_x(f_x)^{b_x} = a_x(\gamma_x m_x P)^{b_x}
    \label{eqn:sol law}
\end{equation}
where $\gamma_x$ is a species fugacity coefficient, $m_x$ is the mol fraction of the species currently in the gas phase, and $P$ is pressure in bars.
$\gamma_{\rm{H_2O}}$ and $\gamma_{\rm{CO_2}}$ are calculated from \citet{holland1991CompensatedRedlichKwongCORKEquation}, $\gamma_{\rm{H_2}}$ from \citet{shaw1964FugacityCoefficientsHydrogen} and all other fugacity coefficients from \citet{shi1992ThermodynamicModeingCHOS}.
Values for $a_x$ and $b_x$ are taken from \citet{burgisser2015SimulatingBehaviorVolatiles}, \ce{CO}, \ce{CH4} and \ce{S2} are treated as insoluble.
Volatile solubility laws are compositionally dependent; for this study all solubility constants were chosen to be suitable for a basaltic melt.
The equilibrium constant equations of eqs.\,\ref{eqn:k1} - \ref{eqn:k5} for each species are solved simultaneously with eq.\,\ref{eqn:sol law}, and mass balance is maintained by keeping the mass of atomic O, C, H and S constant.

We note here that this is a somewhat simplistic model for sulfur, which does not account for how sulfur degassing can impact the \fo{} of a melt as it erupts.
Source-to-surface modelling of a magma degassing \ce{H2} would show sulfur species further complicating the relationship between mantle \fo{} and \ce{H2} outgassing, as release of S$^{2-}$ can lower the \fo{} of a system \citep[e.g.,][]{gaillard2011AtmosphericOxygenationCaused}, while sulfur degassing is enhanced under oxidising conditions.
Given that the \fo{} conditions relevant to most planetary magmatism will have sulfur speciated as S$^{2-}$, the result of our simplified S degassing model is to slightly underestimate the \ce{H2} production, if the \fo{}'s we quote are viewed as being mantle \fo{}.

\subsection{Volcanic outgassing}
The outgassing flux of a species can be parameterized in terms of a concentration of species X as

\begin{equation}
\Phi_{\mathrm{volc}}(X) = \Phi_{\rm volc} \: f^{\mathrm{v}}_{\mathrm{T}}(X),
\end{equation}
where $\Phi_{\rm volc}$ is the total volcanic outgassing flux [mol s$^{-1}$], and $f^{\mathrm{v}}_{\mathrm{T}}(X)$ is the total number mixing ratio of species $X$ in the volcanic input to the base of the atmosphere.

The volcanic outgassing flux varies according to the vent pressure and melt \fo{} (See \ref{appendix_outgass}).
To account for this, Earth's modern flux \citep[$\Phi_{V0}$ = $3.4 \times 10^{6}$ mol\,s$^{-1}$, taken from Tables 7.1 and 7.3 in][]{catling2017AtmosphericEvolutionInhabited}, was assumed to be equal to the amount of volcanic outgassing modelled for a magma degassing at 1\,bar and $\Delta$FMQ=0, with an initial volatile content of 350\,ppm H, 550\,ppm C and 1000\,ppm S \citep[approx 0.3 wt\,\% \ce{H2O} and 2000\,ppm \ce{CO2}; this modern Earth reference run was chosen as a good approximation of the average modern melt compositions degassing at mid-ocean ridges,][]{michael2015BehaviorConcentrationCO2,wallace2015ChapterVolatilesMagmas}.
In all cases, the initial volatile content refers to an undegassed magma, rather than a mantle source.
The number of moles from all other model runs were then scaled against the total number of moles of gas (of all species) released by this reference run for the modern Earth,

\begin{equation}
\Phi_{\mathrm{volc}} = \dfrac{n}{n_{\rm{FMQ}}} \cdot \Phi_{V0},
\end{equation}
where $n$ is the total number of moles of gas released by any one model run, and $n_{\rm{FMQ}}$ is the number of moles released by our reference modern Earth run. 
This ratio is used to adjust the volcanic flux to account for the effect of redox and surface pressure on volcanic degassing.

Volcanic outgassing of hydrogen is discussed here as \ftvh{}, the total hydrogen (H, tracked as \ce{H2} molecules) number mixing ratio of the volcanic gases input to the base of the atmosphere. This value can be parameterized in several different ways,

\begin{subequations}
	\label{eqn:3eqs}
	  \begin{align}[left ={f^{\mathrm{v}}_{\mathrm{T}}(\mathrm{H}_2) \approx \empheqlbrace}]
		& f^{\rm v}(\ce{H_2}),\label{eqn:ft_H2_lower}\\
		& f^{\rm v}(\ce{H_2}) + 4f^{\rm v}(\ce{CH4)} + 3f^{\rm v}(\ce{H2S}),\label{eqn:ft_H2_upper}\\
		& f^{\rm v}(\ce{H_2}) + 2f^{\rm v}(\ce{CH4)} + f^{\rm v}(\ce{H2S}), & \label{eqn:ftv_H2}
	   \end{align}
	  \end{subequations}
where $f^{\rm v}(X)$ is the mixing ratio of each individual species $X$ (i.e., \ce{H2}, \ce{CH4} etc) in the volcanic gas. of The three different cases in eq.\ref{eqn:3eqs}, summarised here, are discussed in more detail in \ref{appendix_param}.
\begin{description}
	\item[Case a:] Only molecular \ce{H2} from volcanic outgassing is considered as providing a \ce{H2} input to the base of the atmosphere. All other hydrogen-bearing species, e.g., \ce{CH4} and \ce{H2S} are assumed to be removed rapidly via deposition to the surface.
	\item[Case b:] All the reduced hydrogen-bearing volcanic gases i.e., \ce{H2}, \ce{CH4} and \ce{H2S} contribute to \ftvh{}. This is equivalent to assuming that all the \ce{CH4} and \ce{H2S} is oxidised to \ce{CO2} and \ce{SO2} respectively, before reaching the homopause, providing an additional volcanically derived flux of \ce{H2} \citep{catling2017AtmosphericEvolutionInhabited}. As \ce{H2O} is the oxidant in both cases, this case requires a large enough reservoir of atmospheric water to oxidise all the \ce{CH4} and \ce{H2S} released, to maintain the photochemically-derived \ce{H2} over the geological timescales of interest.  
	\item[Case c:] Some volcanic \ce{H2S} will be deposited to the surface, and some volcanic \ce{H2} will be removed from the atmosphere as it undergoes subsequent photochemical and thermochemical reactions followed by rainout of any soluble species produced. The rest of the \ce{CH4} and \ce{H2S} reacts as in case b, leaving \ftvh{} in this case to be the sum of the H-bearing species, weighted by the number of moles of \ce{H2} each species contains (equivalent to all the H-bearing species being photodissociated in the upper atmosphere).
\end{description}
In a natural system, it is very unlikely that all the outgassed \ce{CH4} and \ce{H2S} will be efficiently converted to \ce{H2} with no deposition to the surface.
We therefore adopt the definition of case c here, with \ftvh{} defined in equation \ref{eqn:ftv_H2}.
This provides a compromise that is intermediate between the definitions of equations \ref{eqn:ft_H2_lower} and \ref{eqn:ft_H2_upper} (See \ref{appendix_param} for examples of how the choice of photochemical regime impacts results). 

\hflux{}, the total flux of molecular hydrogen from volcanic sources, can therefore be calculated as:

\begin{equation}
\Phi_{\mathrm{volc}}(\ce{H2}) = \Phi_{\rm volc} \cdot f^{\rm v}_\mathrm{T}(\ce{H_2}) = \Phi_{\rm volc} \: (f^{\rm v}(\ce{H_2}) + 2f^{\rm v}(\ce{CH4)} + f^{\rm v}(\ce{H2S})).
\end{equation}
In order to consider atmospheric escape, it is convenient to convert from geological units of mol\,s$^{-1}$ to the photochemical units molecules\,cm$^{-2}$\,s$^{-1}$.

Although volcanic CO can generate \ce{H2} through

\begin{equation}
	\ce{CO} + \ce{H2O} \leftrightharpoons \ce{CO2} + \ce{H2},
\end{equation}
we treat volcanic CO outgassing as compensated for by CO deposition, i.e.,

\begin{equation}
	\Phi_{\mathrm{volc}}(\ce{CO}) = v_{\mathrm{dep}}(\ce{CO})f(\ce{CO}),
\end{equation}
as the deposition rate is expected to overwhelm the photochemistry. $\Phi_{\mathrm{volc}}(\ce{CO})$ is the volcanic flux of \ce{CO}, $v_{\mathrm{dep}}(\ce{CO})$ the deposition velocity [cm s$^{-1}$] and $f(\ce{CO})$ the atmospheric mixing ratio of \ce{CO}.

\section{Relating outgassing to atmospheric \ce{H2} mixing ratios}

We initially assume the rate of escape of \ce{H2} to space is limited by the diffusion rate across the homopause, as is the case on the modern Earth. The diffusion limited flux is given by \citet{walker1977EvolutionAtmosphere} as

\begin{equation}
	\label{eqn:diff_lim}
	\Phi_{\ce {esc}}(\ce{X}) \cong \frac{b_{\ce{X}}}{H_a} \cdot \frac{f_T(\ce{X})}{1 + f_T(\ce{X})}
	  \end{equation}
where b$_{\ce{X}}$ is the binary diffusion parameter for species X, $H_a$ is the scale height of the atmosphere, and $f_T(\ce{X})$ is the total number mixing ratio of X in the atmosphere (as opposed to $f^{\mathrm{v}}_{\mathrm{T}}(X)$, which is the mixing ratio in the volcanic gas).

$f_T(\ce{X})$ here is \fth{}, the sum of the mixing ratios of all \ce{H2}-bearing atmospheric constituents above the tropopause \citep[e.g.,][]{kharecha2005CoupledAtmosphereEcosystem}. \fth{} is evaluated above the stratospheric cold trap, where \ce{H2O} has been removed by condensation,

\begin{equation}
    f_{\mathrm{T}}(\mathrm{H}_2) \approx f(\ce{H_2}) + 2f(\ce{CH4)} + f(\ce{H2S}).
\label{eqn:ft_H2}
\end{equation}
We use a well established convention for the escape of \ce{H2} \citep{walker1977EvolutionAtmosphere, kasting1993EarthEarlyAtmosphere,kharecha2005CoupledAtmosphereEcosystem}, by balancing the volcanic outgassing flux $\Phi_{\mathrm{volc}}$ with escape, i.e.,

\begin{equation}
    \Phi_{\mathrm{volc}}(\ce{H2}) \cong \Phi_{\ce {esc}}(\ce{H2}) = \alpha \cdot \Phi_{\ce {diff}}(\ce{H2}).
\end{equation}
The escape flux $\Phi_{\ce {esc}}$ can be less than or equal to the diffusion limited case $\Phi_{\ce {diff}}$, with the difference parameterized by an escape efficiency factor $\alpha$ bound between 0 and 1.

Volcanic flux is also a variable, so this is examined using an outgassing flux factor $\beta$, indicating the factor of increase in volcanic outgassing rate compared to the modern Earth. This maintains a steady-state atmosphere, with the total fraction of \ce{H2} in the atmosphere defined as

\begin{equation}
    \frac{f_{\mathrm{T}}(\mathrm{H}_2)}{1 + f_{\mathrm{T}}(\mathrm{H}_2)} = \dfrac{\beta \: \Phi_{\mathrm{volc}}(\ce{H2}) \: H_a}{\alpha \: b_{H_n}}.
\label{eqn:fth2}
\end{equation}
$b_{H_n}$ is the binary diffusion parameter at  1.8$\times10^{19}$\,\percmpers, the weighted average of $b_{\ce{H}}$ (2.73$\times 10^{19}$\,\percmpers) and $b_{\ce{H_2}}$ (1.46$\times 10^{19}$ \percmpers) based on the modern relative abundances of H and \ce{H2} at the homopause.
In detail, these values will vary slightly depending on the homopause temperature and background atmospheric composition.
To find the reduction in escape efficiency necessary to achieve a given \fth{}, we solve for $\alpha$

\begin{equation}
    \alpha = \dfrac{\beta \: \Phi_{\mathrm{volc}}(\ce{H2}) \: H_a}{b_{H_n}} \cdot \dfrac{1 + f_{\mathrm{T}}(\mathrm{H}_2)}{f_{\mathrm{T}}(\mathrm{H}_2)},
\label{eqn:alpha}
\end{equation}
setting \fth{} to a desired atmospheric mixing ratio of \ce{H2} and keeping in mind that $\alpha$ is an efficiency factor bound between 0 and 1, defining how far below the diffusion limited loss rate \ce{H2} escape is occurring.

\section{\fth{} systematics with \fo{} and pressure}
We initially examine how \fth{} varies with \fo{} and vent pressure alone, across three different initial volatile contents.
Different pressures are considered to cover the range of vent pressures relevant for the Archean \citep[from 3 to $<$0.5\,bar, e.g.,][]{goldblatt2009NitrogenenhancedGreenhouseWarming,som2016EarthAirPressure,rimmer2019OxidisedMicrometeoritesEvidence}.
They also cover pressures relevant to vents in submarine settings under shallow oceans, a possible scenario for the Archean \citep{flament2008CaseLateArchaeanContinental}.

\begin{figure}[H]
\begin{center}
\includegraphics[width=0.8\textwidth]{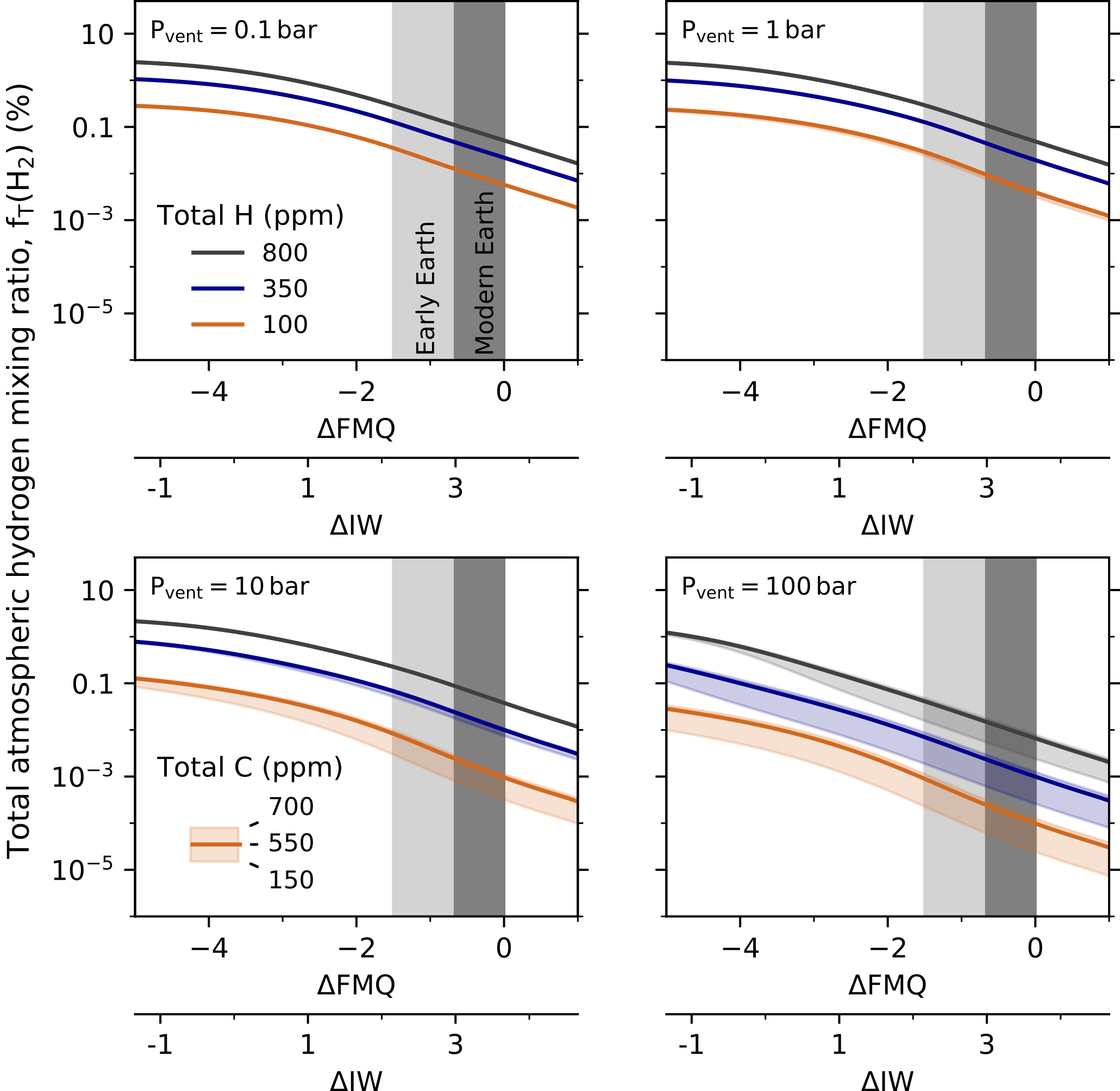}
\caption{Total hydrogen mixing ratio sustained in the troposphere according to the oxygen fugacity of the degassing magma. Calculations assume the loss of hydrogen to space is diffusion limited ($\alpha$=1), and consider four different vent pressures all with a volcanic flux equal to the modern Earth ($\beta=1$). Results are shown for three different H contents, all with 1000\,ppm S and, and indicating the effect of a variable C content between bounds of 150 and 700\,ppm C. At FMQ and 1\,bar pressure, 150, 550 and 700\,ppm C correspond to approx 550, 2000 and 2500\,ppm total \ce{CO2} in the system, and 0.09, 0.3 and 0.7 wt\,\% \ce{H2O} for 100, 350, 800\,ppm H respectively. Both the \ce{CO2} and \ce{H2O} contents are almost entirely partitioned into the gas phase at 1\,bar pressure. The modern range in MORB \fo{} is highlighted as a dark grey bar \citep{bezos2005Fe3SFeRatios}, with a possible lower bound on Archean \fo{} represented in light grey \citep{aulbach2016EvidenceReducingArchean, nicklas2019SecularMantleOxidation}.}
\label{fig:mixing ratio}
\end{center}
\end{figure}

Depending on the initial volatile content of the melt (see \ref{appendix_outgass}) and the vent pressure, more reduced melts can sustain an atmosphere with a greater fraction of \ce{H2} (Fig. \ref{fig:mixing ratio}), with a decrease in \fo{} of 2 log units providing a 15$\times$ increase in \fth{}.
\fth{} converges on a maximum value at low \fo{}, which is dependent on greatest achievable \ftvh{} for a melt's volatile content and vent pressure (see \ref{appendix_outgass}).
\fth{} for a fixed \fo{} decreases with increasing surface pressure, although there is very little difference in the results for 0.1 and 1\,bar.
This is due to the pressure dependent solubility of water and sulfur species; at high vent pressures a smaller fraction of the system's volatiles have exsolved and are available to form the \ftvh{} fraction, effectively decreasing the volcanic flux.
Atmospheres fed by H-poor magmas show a greater decrease in \fth{} as surface pressure is increased.
The scale of this decrease is greater the more oxidised the system is.
H-poor systems are particularly sensitive, as water is very soluble in basalt: for a 100\,ppm H system at \fo{}$\,>\,$FMQ\,-3, 45\,\% of the H content (almost entirely speciated as water) can remain dissolved in the melt even at 1\,bar pressure.
In contrast, for an 800\,ppm H system, $<$\,10\,\% of the total hydrogen is still stored in the melt at 1\,bar and with an \fo{}$\,>\,$FMQ\,-1.

Higher pressure also results in greater variability of \fth{} with carbon content. At low pressures, and to some extent low \fo{}, a carbon content change of $\sim$4.5$\times$ produces a negligible impact on atmospheric \fth{}. However, at high pressures and higher \fo{}, this range changes the \fth{} by almost 1 order of magnitude. \fth{} increases with C content at high pressures because the magma reaches \ce{CO2} saturation earlier, and forms a gas phase within the melt at higher pressures. Hydrogen bearing species can then partition into this gas phase and be degassed at higher pressures than they would be in a C-poor magma.

\section{Achieving volcanically-sustained \ce{H2} atmospheres}

Fig.\,\ref{fig:mixing ratio} suggests that for a very reduced planet with a moderate to low pressure atmosphere and with a mantle volatile budget, volcanic flux, and stellar environment otherwise similar to the modern Earth, the maximum \fth{} achievable is 3.2\,\%.
More conservatively, considering the range of magma \fo{} likely across Earth history \citep[e.g.,][]{aulbach2016EvidenceReducingArchean,nicklas2019SecularMantleOxidation,bezos2005Fe3SFeRatios}, represented as the dark and light grey bars in Fig.\,\ref{fig:mixing ratio}, the maximum possible \fth{} from this model would be 0.4\,\%, assuming $\leq$ 1\,bar surface pressure and a high H content.
This result is consistent with the 0.1\,\% \fth{} often quoted for the Archean \citep{walker1977EvolutionAtmosphere, kasting1993EarthEarlyAtmosphere}, but it requires degassing pressures of less than 10\,bar and $>$350\,ppm H in the melt.
To achieve values of \fth{} more conducive to greenhouse heating, prebiotic chemistry, and to match evidence for $\sim$\,1\,\% \fth{} \citep{zahnle2019StrangeMessengerNew}, an increase in the volcanic flux ($\beta >1$, Fig.\,\ref{fig:flux_only}), or a reduction in the loss efficiency, $\alpha < 1$ (Fig.\,\ref{fig:alpha}) must be invoked. We therefore explore varying these two parameters in this section.

\subsection{Changing the volcanic flux}

\begin{figure}[ht]
	\begin{center}
	\includegraphics[width=0.9\textwidth]{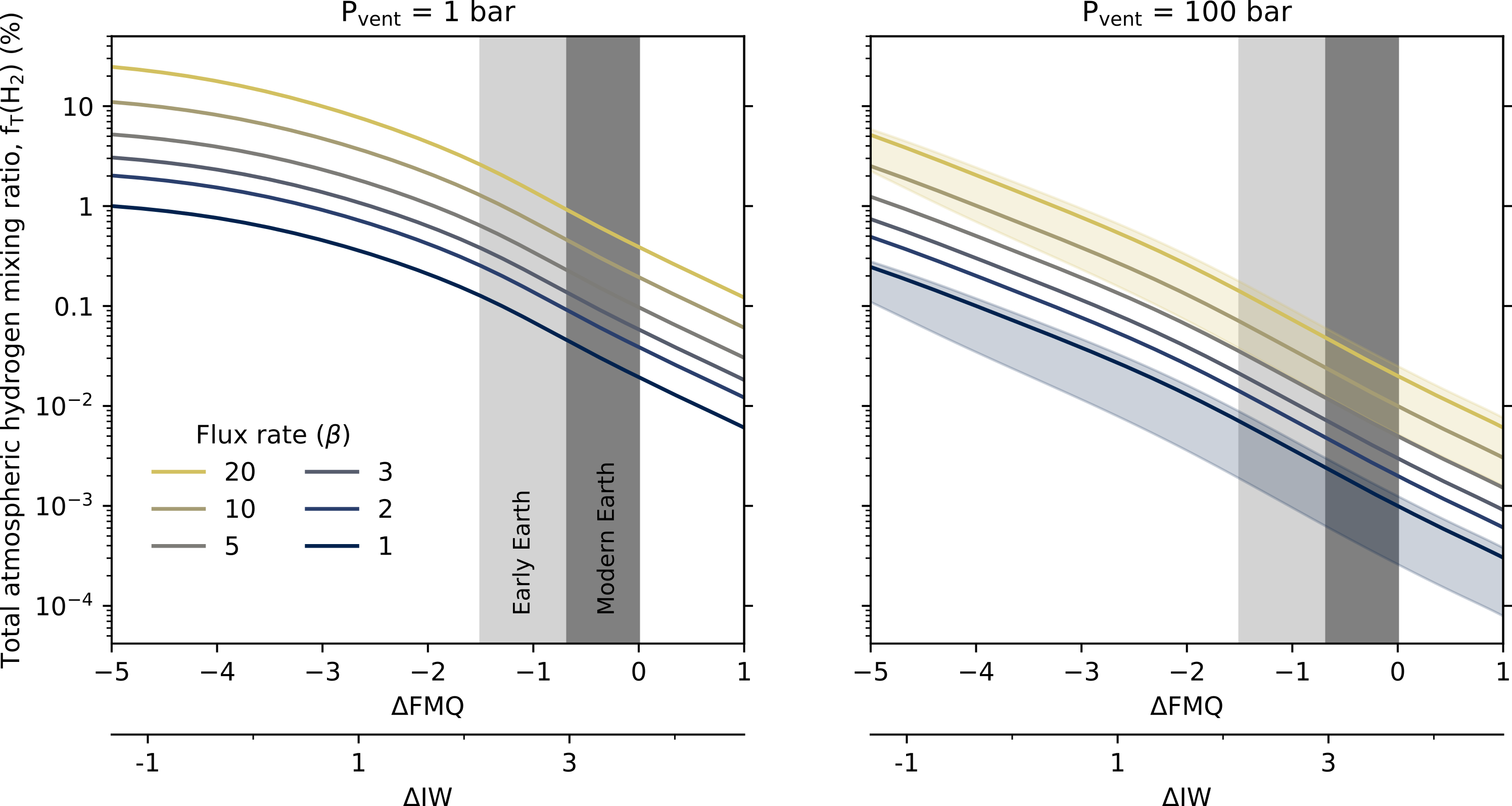}
	\caption{The effect of a higher volcanic flux on the hydrogen mixing ratio, \fth{}, in the atmosphere.
	Calculated with a volatile content of 350\,ppm H, 550\,ppm C and 1000\,ppm S at 1 and 100\,bar, $\alpha$=1.
	Lines are coloured according to the flux, and the dark/light grey bars represent the \fo{} of modern Earth MORB and possible early Earth basalts respectively. Shaded bands of uncertainty in the 100\,bar panel represent the area of 150 - 700\,ppm C in the melt for flux rates of 1 and 20$\times$.}
	\label{fig:flux_only}
	\end{center}
\end{figure}

A change in the volcanic flux can modify the \ce{H2} flux to an atmosphere (see eq.\,\ref{eqn:fth2}, Fig. \ref{fig:flux_only}) and is representative of either, 1) a change in the magma's initial volatile content, 2) a greater proportion of magma reaching low pressure, or 3) increased magma production.
Here, we have examined a range of volcanic fluxes from modern to 20$\times$ the modern rate.
As expected, increasing the volcanic flux produces a nearly proportional increase in \fth{} (Fig.\,\ref{fig:flux_only}); proportional where \fth{} $\ll$ 1, and slightly less than a proportional as \fth{} increases above this point.
Increasing the H content from 350\,ppm to 800\,ppm is equivalent to increasing the volcanic flux by 2-3$\times$.
For a magma at FMQ\,-1.5, increasing the volcanic flux to 10$\times$ modern produces an \fth{} of 1.6\,-\,3.4\,\% for 350\,ppm H and 800\,ppm H, respectively.

The 100\,bar panel in Fig. \ref{fig:flux_only} indicates that even with very high fluxes and a reduced Archean melt, the maximum achievable \fth{} would be around 0.2\,\%.
Only a highly reduced planet erupting melts around the IW buffer, with a high volatile content (equivalent to $>$400\,ppm H) would see \fth{} contents above 1\,\% in an atmosphere if the average vent pressure is at 100\,bar.
We therefore suggest that without significant modification to the hydrogen escape rate, it is unlikely than an early Earth scenario with the majority of it's volcanism occurring in submarine settings could attain a hydrogen-rich secondary atmosphere. Exoplanets covered by global oceans, and those with dense atmospheres (e.g., \ce{CO2}-rich atmospheres similar to that of Venus) would similarly have to be erupting highly reduced melts with a high volcanic flux, and/or have a lower hydrogen escape rate to achieve an atmosphere with an \fth{} on the order of 1\% or above.

\subsection{Changing the \ce{H2} escape rate}

\begin{figure}[!ht]
	\begin{center}
	\includegraphics[width=0.9\textwidth]{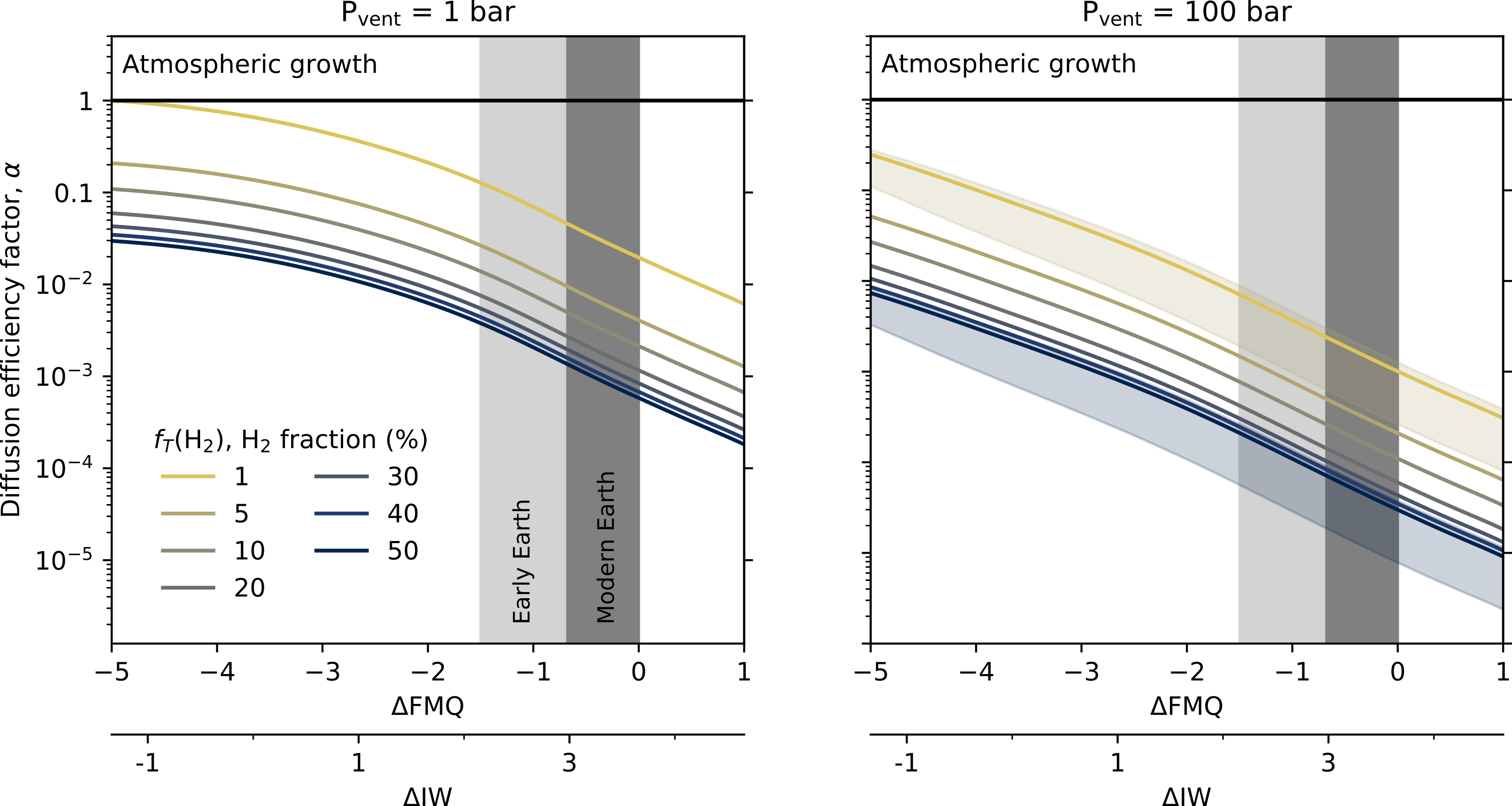}
	\caption{Values of alpha required to achieve a certain hydrogen mixing ratio in the atmosphere, dependent on the \fo{} of the degassing melt. For melt with 350\,ppm H, 550\,ppm C, 1000\,ppm S, degassing at either 1 or 100 bar vent pressure, $\beta=1$. Shaded bands of uncertainty in the 100\,bar panel represent the area of 150 - 700\,ppm C in the melt for \fth{} = 1 and \fth{} = 50. An $\alpha$ value $>1$ indicates a diffusion limited loss rate greater than today's would be necessary to limit the \ce{H2} content of the atmosphere to the specified fraction; as this is unlikely, we infer that \fth{} would increase until steady state is reached at $\alpha = 1$.}
	\label{fig:alpha}
	\end{center}
\end{figure}

Rather than modifying the volcanic input term to the atmosphere, hydrogen can be built up in the atmosphere by reducing the loss rate to space.
So far, we have considered loss rate to be purely diffusion-limited, as it is on the modern Earth \citep{catling2017AtmosphericEvolutionInhabited}.
However, if the loss rate were less efficient, a greater \fth{} could be achieved for the same \fo{} and volcanic flux. We have parameterized the reduction in loss efficiency as $\alpha$, and calculated the value of $\alpha$ required to achieve a certain \fth{}, for a given magma \fo{} and $\beta$=1 (Fig. \ref{fig:alpha}).

At the reduced end of Fig. \ref{fig:alpha}, around IW, a loss rate of $\alpha$\,=\,0.1 (i.e., 10\,\% of modern) is enough to build up an \fth{} of 10\,\%.
In contrast, across the more oxidised magmatic conditions prevailing over Earth history, an $\alpha$ of 0.16 - 0.025 is needed to build up a 1\,\% mixing ratio of \ce{H2} in a 1\,bar atmosphere.
Increasing the volatile content to 800\,ppm H allows $\alpha$ to increase to 0.4-0.06 while still generating 1\,\% \ce{H2} at 1\,bar.
However, once the vent pressure has been increased to 100\,bar, $\alpha < 8\times 10^{-3}$ is required to achieve 1\,\% \fth{} given Earth's magmatic \fo{}.
Large reductions in escape efficiency below the diffusion limit are required to obtain \fth{} at the percent level.
We discuss whether such values for $\alpha$ are plausible in the following section, but this suggests that the volatile content and rate of volcanism could be the key variables in achieving a hydrogen-rich secondary atmosphere, rather than large changes in the atmospheric loss rate.

\section{\ce{H2} on the early Earth}

Using a thermodynamic magma degassing model, we have mapped the parameter space for producing a terrestrial atmosphere with a significant \ce{H2} mixing ratio.
The parameters we have considered are magma \fo{}, volatile content, volcanic outgassing flux, and atmospheric escape efficiency as variables.
A 6 log unit change in \fo{} can produce a change in \fth{} of around a factor of 100.
On planetary bodies with a reduced mantle such as Mars, volcanic fluxes comparable to those likely over the evolution of the Earth \citep[1-12$\times$ modern e.g.,][]{korenaga2006ArcheanGeodynamicsThermal, sleep2001CarbonDioxideCycling, avice2017OriginDegassingHistory} could result in secondary atmospheres with hydrogen fractions on the order of 1-10's of percent.
However, over the restricted range of \fo{} likely for the Archean Earth, the most effective way to achieve an \fth{} $>$ 1\% is to vary both the outgassing flux and the loss efficiency within ranges closer to modern.

\subsection{Likely escape rates on the early Earth}

The escape rate of hydrogen from an atmosphere with a given \fth{} will decrease below the diffusion limited escape rate, if removal of \ce{H2} from above the homopause becomes less efficient.
In this scenario, \ce{H2} escape becomes energy limited by the incident XUV flux \citep[e.g.,][]{tian2005HydrogenRichEarlyEarth}. \cite{luger2015ExtremeWaterLoss} describe the energy-limited mass loss rate ($\dot{m}_{\rm XUV}$) as

\begin{equation}
	\dot{m}_{\rm XUV} = \dfrac{\epsilon_{\rm XUV} \pi \mathcal{F}_{\rm XUV} R_{\rm p} R_{\rm XUV}^{2}}{GM_{\rm p} K_{\rm tide}} \qquad \rm (g\,s^{-1}),
\end{equation}
where $\epsilon_{\rm XUV}$ is the absorption efficiency of XUV, $\mathcal{F}_{\rm XUV}$ is the XUV flux at the top of atmosphere, $M_{\rm p}$ is the mass of the planet, $R_{\rm p}$ is the planet radius, $R_{\rm XUV}$ is the radius where the bulk of the energy is deposited (which, for simplicity, we take to be equal to $R_{\rm p}$), and $K_{\rm tide}$ is a tidal correction term of order unity.
We convert this equation into units of flux,

\begin{equation}
	\Phi_{\rm XUV} = \dfrac{\epsilon_{\rm XUV} S \mathcal{F}_{\rm XUV} R_{\rm p} N_{A}}{4GM_{\rm p}M_{\ce{H2}}} \qquad \ce{H2}\,\rm (molecules\,cm^{-2}\,s^{-1}),
\label{eqn:xuv_flux}	
\end{equation}
where $S$ is the XUV irradiation relative to modern Earth, $\mathcal{F}_{\rm XUV}$ is the modern XUV flux of 4.5 ergs\,\percmsqpers{} for Earth \citep{ribas2005EvolutionSolarActivity},  $N_{A}$ is Avogadro's constant ($6.022\times10^{23}$\,mol$^{-1}$) and $M_{\ce{H2}}$ is the molecular mass of a \ce{H2} molecule ($2.02$\,g\,mol$^{-1}$).
Eq.\,\ref{eqn:xuv_flux} shows that as planetary mass decreases, the $\Phi_{\rm XUV}$ increases, getting closer to the diffusion limited rate ($\Phi_{\rm diff}$).
At the same time, the diffusion limited flux decreases as $\Phi_{\rm diff} \propto$ gravitational strength. Hence as planetary mass decreases, the difference between the energy limited and diffusion limited fluxes decreases, and the resulting \ce{H2} escape flux is reduced. Eq.\,\ref{eqn:xuv_flux} is also highly dependent on the absorption efficiency of XUV, $\epsilon_{\rm XUV}$. This value is typically fixed at 0.1 for terrestrial planets with high molecular weight atmospheres \citep[e.g.,][]{lopez2012HOWTHERMALEVOLUTION, owen2015UVDRIVENEVAPORATION, bourrier2017TemporalEvolutionHighenergy}, and is linked to the availability of different atmospheric species. For example, high \ce{H2} fractions cause more efficient XUV absorption, a greater heating rate and faster escape; the presence of IR cooling molecules such as \ce{CO2} reduces the amount of thermal energy which can be converted into atmospheric escape, reducing the loss rate.

\begin{figure}[H]
	\begin{center}
	\includegraphics[width=\textwidth]{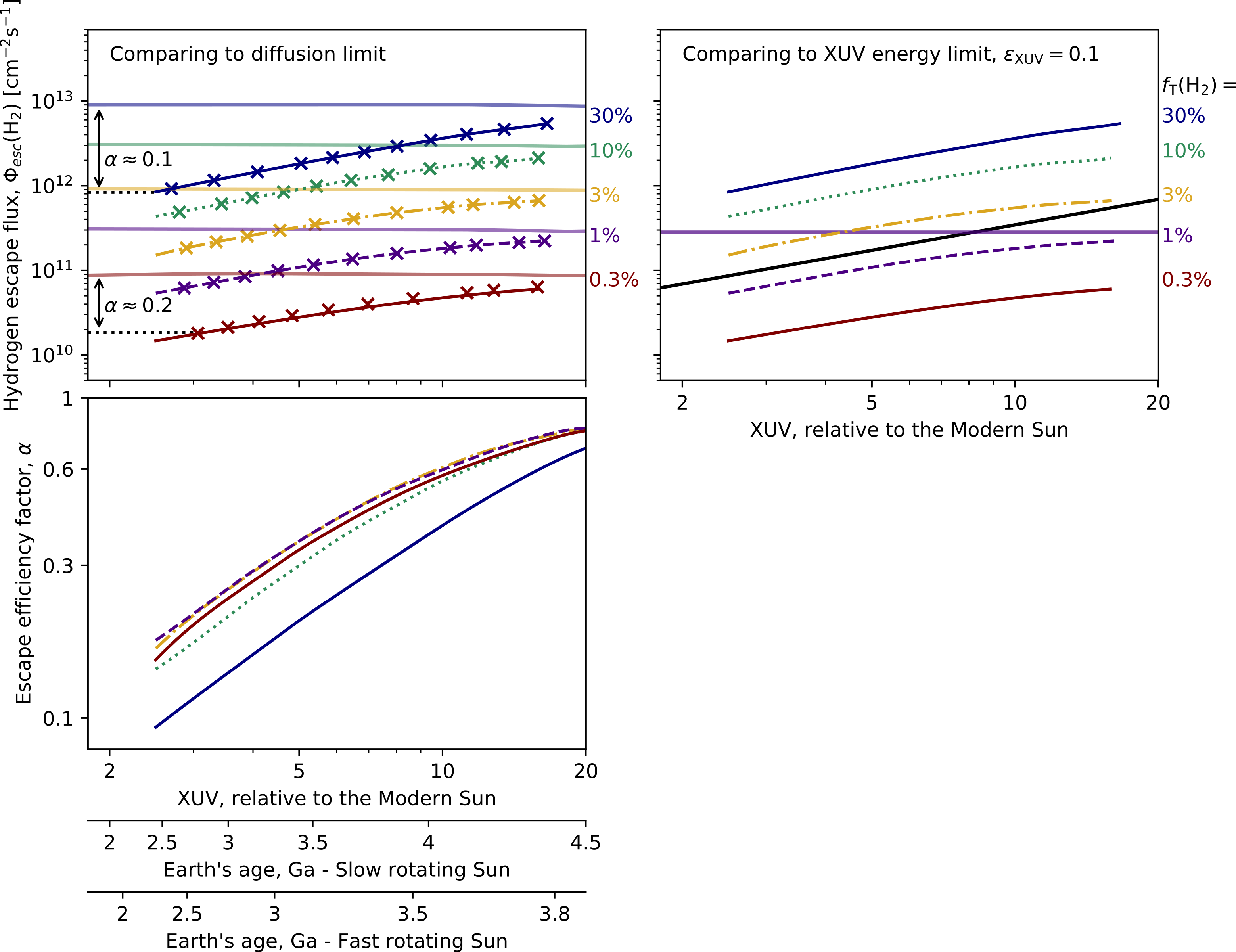}
	\caption{Top left: Crossed lines represent the hydrodynamic hydrogen flux from an atmosphere with an exobase temperature of 250\,K, with the solid lines representing the diffusion limited flux for the same atmosphere. Escape fluxes have been calculated for several different values of \fth{} \citep[taken from][]{zahnle2019StrangeMessengerNew}, represented by changing colour and dash type, as a function of XUV flux (Extreme UV).
	Top right: The same hydrodynamic fluxes are shown, this time compared to an XUV limited flux (diagonal black line) calculated using Eq.\ref{eqn:xuv_flux}, $\epsilon_{\rm XUV} = 0.1$ and the diffusion limit for 1\,\% \fth{} (horizontal line) as a reference.
	Bottom: The difference between the diffusion limited flux and the hydrodynamic loss rate gives $\alpha$ ($\alpha$=$\frac{\text{hydrodynamic loss rate}}{\text{diffusion lim. loss rate}}$), which is primarily a function of XUV flux, and largely insensitive to \fth{} for most values below 30\,\%. An $\alpha$ of 1 is an escape efficiency equal to the diffusion limited escape rate.}
	\label{fig:xuv}
	\end{center}
\end{figure}

Modelling of the hydrogen escape flux compared to diffusion limited and energy limited escape rates, depending on the activity of the Archean sun, has been carried out by \cite{tian2005HydrogenRichEarlyEarth,kuramoto2013EffectiveHydrodynamicHydrogen} and \cite{zahnle2019StrangeMessengerNew} (with these results summarised in Fig.\,\ref{fig:xuv}).
These models of the early Earth use a very low temperature for the exobase (250 K compared to the modern 1000 K), after assuming an anoxic and \ce{CO2}-rich atmosphere, similar to Mars and Venus today.
A cold exobase makes the Jeans' escape rate minimal, so that hydrodynamic escape controls escape efficiency.
However, the temperature of the Archean upper atmosphere is highly debated, and will strongly depend on the atmospheric composition \citep{catling2006CommentHydrogenRichEarly} and non-LTE effects, the operation of which in the Archean atmosphere are poorly constrained.
A warmer exobase would increase the overall rate of loss, making it more likely the loss occurred at, or close to, the diffusion limit; therefore the hydrodynamic loss rates seen below should be considered a lower limit of escape rate and $\alpha$.

In Fig. \ref{fig:xuv} the escape flux from the model of \cite{zahnle2019StrangeMessengerNew} is compared to both the diffusion limited escape rate, and an XUV energy-limited rate.
Young stars are more active, and therefore emit more XUV radiation earlier in their life-cycles, so the average incident XUV ($\mathcal{F}_{\rm XUV}$) at 2.5\,Ga is thought to be around 2.5-3$\times$ today's, up to 5 -10$\times$ modern at 3.5\,Ga depending on whether the Sun was a slow or fast rotator early in it's lifetime \citep{tu2015ExtremeUltravioletXray}.
Higher radiation levels drive higher rates of hydrogen loss above the exobase following the slope of XUV limited escape, until loss starts to asymptote to the diffusion limit, i.e., is limited by supply of \ce{H2} to the hydrogen escape region.
During the solar cycle today, the XUV flux can vary by up to a factor of 5 between a quiet and active sun; however, this variation occurs on a high frequency ~11 year cycle, so has been ignored in preference for an average XUV flux to calculate \fth{}.
$\alpha$ is defined as

\begin{equation}
	\alpha = \frac{\text{hydrodynamic loss rate}}{\text{diffusion limited loss rate}},
\end{equation}
which is plotted in Fig.\,\ref{fig:xuv}, representing a fractional reduction in the escape efficiency from the diffusion limited case.
Fig. \ref{fig:xuv} suggests that a reasonable range for $\alpha$ on the early Earth, assuming an \fth{} of 10\,\% or less, is $\sim$0.2 at 2.5\,Ga, and 0.5-0.6 at 3.5\,Ga. Comparing these values of $\alpha$ to those explored in Fig. \ref{fig:alpha}, suggest that achieving an \fth{} on the order of 1\,\% or higher seems implausible for a high pressure degassing scenario on the early Earth, and indicates that an increase in the volcanic flux compared to modern is likely to be necessary even at lower degassing pressures.

\subsection{Hydrogen mixing ratios on the Archean Earth}

\begin{figure}[H]
	\begin{center}
	\captionsetup[subfigure]{justification=centering}
	\begin{subfigure}[t]{.65\textwidth}
	  \centering
	  \includegraphics[width=\linewidth]{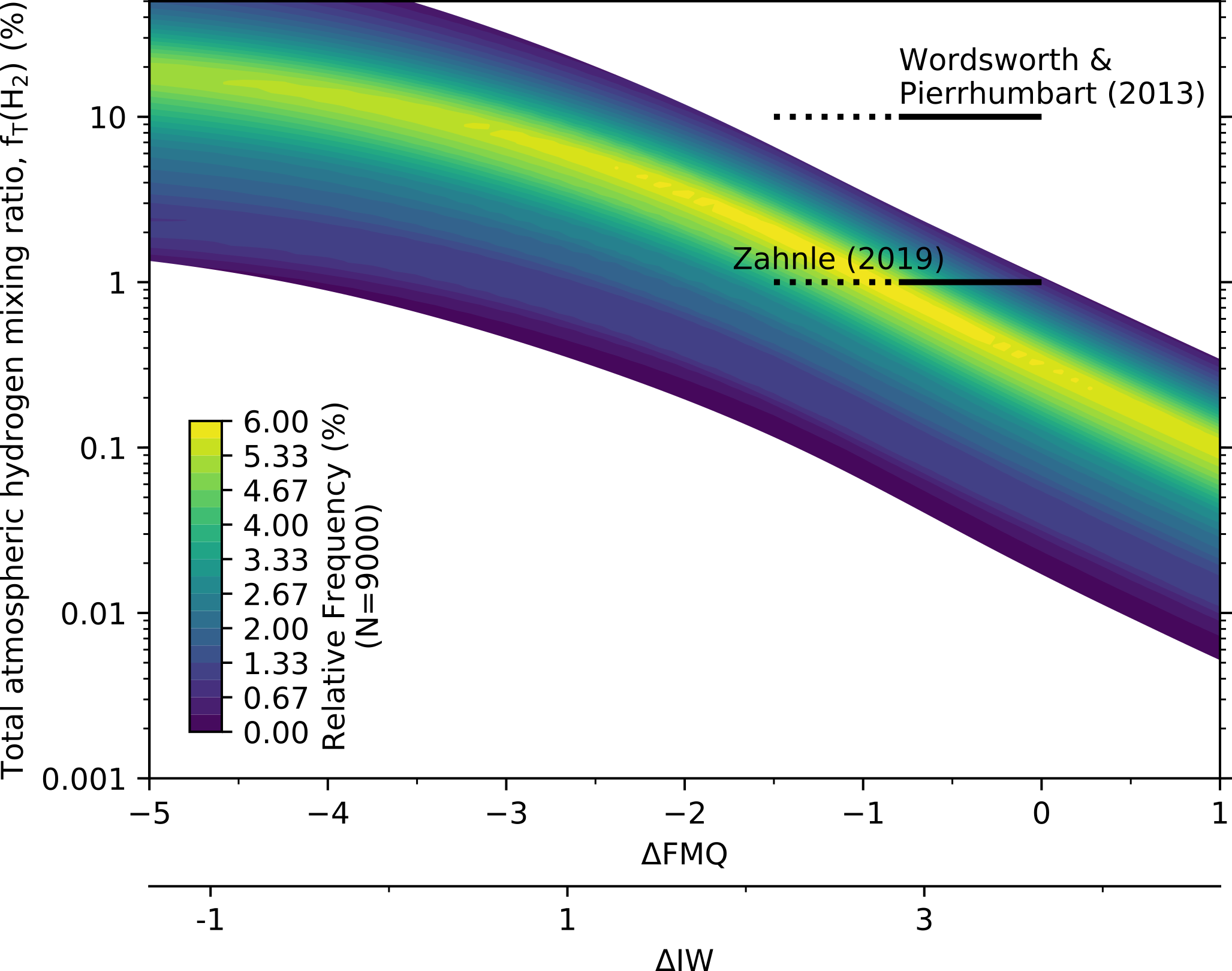}
	  \caption{} 
	  \label{fig:prob_e_cont}
	\end{subfigure}\hfill%
	\begin{subfigure}[t]{.24\textwidth}
	  \centering
	  \includegraphics[width=\linewidth]{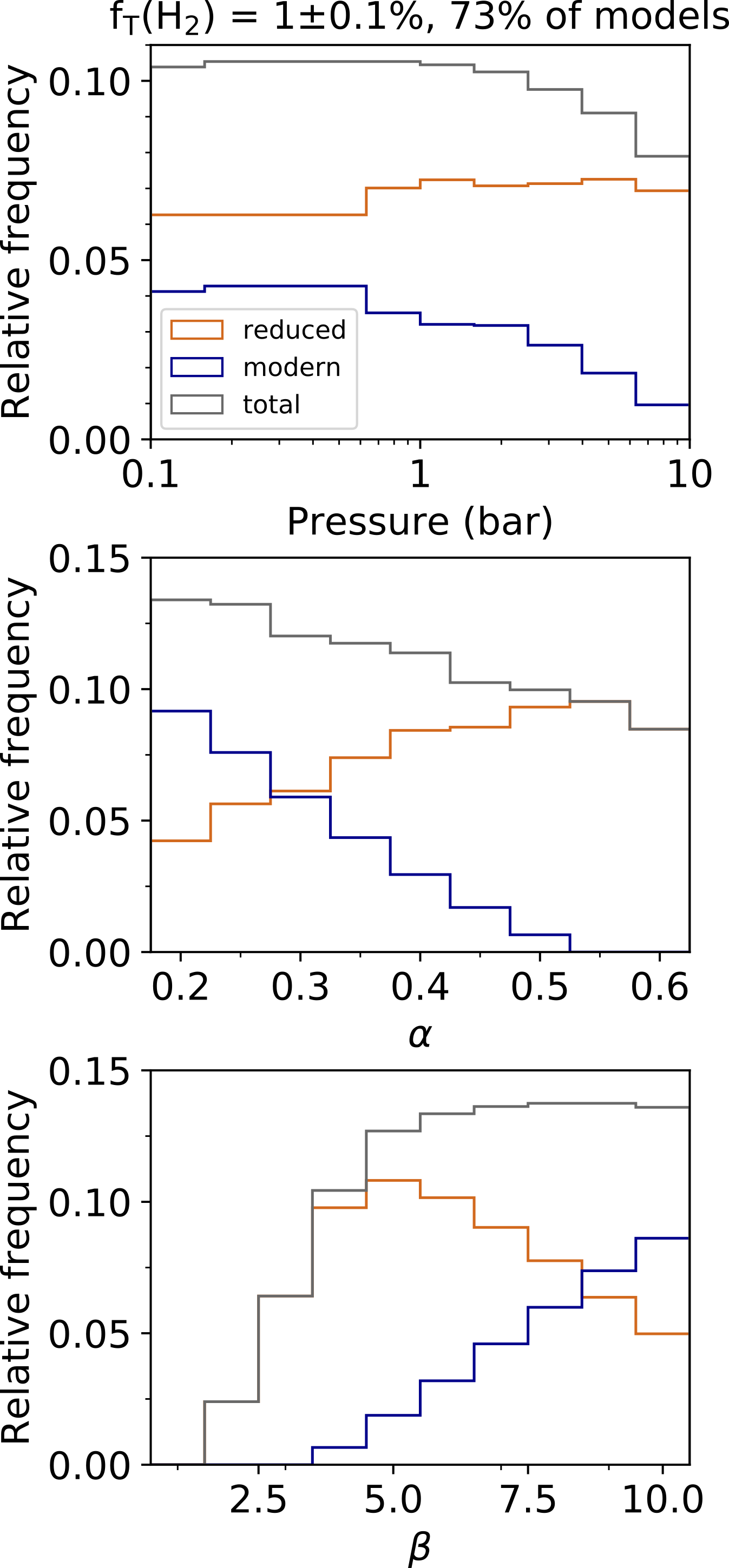}
	  \caption{} 
	  \label{fig:priors}
	\end{subfigure}%
	\caption{(a) The relative frequency of hydrogen mixing ratios in an atmosphere as a function of the magma \fo{}, represented as a 2D density surface. This calculation considers varying parameters uniformly over the ranges, $\alpha$ = 0.2-0.6, $\beta$ = 1 - 10 and surface pressures of 0.1 - 10 bar.
	White regions are unphysical.
	Horizontal bars indicate 1 and 10\,\% \fth{}, plotted across the modern Earth \citep[solid bar][]{bezos2005Fe3SFeRatios} and early Earth \citep[dashed bar]{aulbach2016EvidenceReducingArchean, nicklas2019SecularMantleOxidation} mantle \fo{} range.
	(b) Relative frequency distributions for the three parameters varied in panel (a), for models which can achieve 1$\pm$0.1\,\% \fth{} within the magmatic \fo{} range of Earth, labelled with the percentage of model runs which meet these constraints.
	The plots show a breakdown of the total distribution between models which calculate 1\,\% \fth{} within the modern (blue) or early Earth (red) \fo{} ranges (grey is total).}
	\label{fig:prob_e}
\end{center}	
\end{figure}

In the previous sections, we have demonstrated that varying both the volcanic flux and the atmospheric loss rate can result in significant variation in atmospheric \fth{} at a given magmatic \fo{}, compared to \fth{} calculated with a modern volcanic flux and $\alpha=1$.
Here, we explore the range of potentially achievable \ce{H2} mixing ratios for a planet with similar stellar and volcanic flux parameters to those suggested for the early Earth, examined over a wide range of \fo{} (Fig.\,\ref{fig:prob_e}).

Fig.\,\ref{fig:prob_e} suggests that for a single value of magma \fo{}, varying surface pressure, volcanic flux and loss rate over plausible ranges can result in a change in the mixing ratio by around a factor of 100.
Assuming an Archean magma \fo{} between FMQ\,-1.5 and FMQ, a 2.5-10$\times$ increase in solar XUV (and therefore a range for $\alpha$ of 0.2-0.6 for an atmosphere with less than 30\,\% \fth{}), a volcanic flux of 1-10$\times$ present and a vent pressure of 0.1-10\,bar, the range of \ce{H2} mixing ratios predicted for the Archean Earth is $\sim$ 0.02 - 7\,\% \ce{H2}, with most models falling in the range 0.2 to 3\,\% \fth{} (0.18 - 4.5\,\% accounting for photochemical uncertainty, see \ref{appendix_param}).

Results from \cite{zahnle2019StrangeMessengerNew} have suggested \ce{H2} mixing ratios on the early Earth of $\sim$1\,\%.
Our parameter sweep shows that this can be achieved with a magmatic \fo{} within the range proposed for the Archean, and with only a weak dependence on degassing pressure and a moderate dependence on volcanic flux and loss rate (Fig.\,\ref{fig:priors}).
Our results also show the trade offs between parameters which achieve this fit to \fth{} = 1\,\%.
Assuming an \fo{} similar to modern, the number of model runs which achieve 1\,\% \fth{} decreases with increasing pressure, strongly decreases as $\alpha$ increases with no runs above $\alpha\,=\,0.5$, and increases when $\beta > 3$; no model runs within the modern \fo{} range meet \fth{}=1\,\% using the modern value of $\beta$.
For a more reduced melt, the trends are more or less reversed, with a peak in model frequencies at around $\beta=$\,5.
The peak in the frequency distribution (Fig.\,\ref{fig:prob_e}) falls between FMQ\,-0.8 and FMQ\,-1, with a bias to the reduced \fo{} range.
Our model indicates that 1\,\% \fth{}, which \cite{zahnle2019StrangeMessengerNew} posit Xenon isotope fractionation is indicative of, is easily achievable for the early Earth (Fig.\,\ref{fig:prob_e}) with 73\,\% of models obtaining 1\,\% \fth{} within Earth's mantle \fo{} range.

\cite{wordsworth2013HydrogenNitrogenGreenhouseWarming} suggest that for an Archean atmosphere with \ce{CO2} levels similar to modern, 10\,\% \ce{H2} and 3$\times$ modern \ce{N2} would be needed to raise the surface temperature above freezing, given the faint young sun.
However within our parameter ranges, 10\,\% \fth{} cannot be achieved (Fig.\,\ref{fig:prob_e}).
We therefore suggest that without alternative sources of \ce{H2}, or a large increase in the \ce{N2} and \ce{CO2}/\ce{CH4} atmospheric partial pressures compared to modern, it is unlikely that sufficient \ce{H2}-enhanced greenhouse warming was occurring to raise Archean surface temperatures above freezing.

\subsection{Alternative \ce{H2} sources on the early Earth}

\begin{figure}[H]
	\begin{center}
	\includegraphics[width=0.7\textwidth]{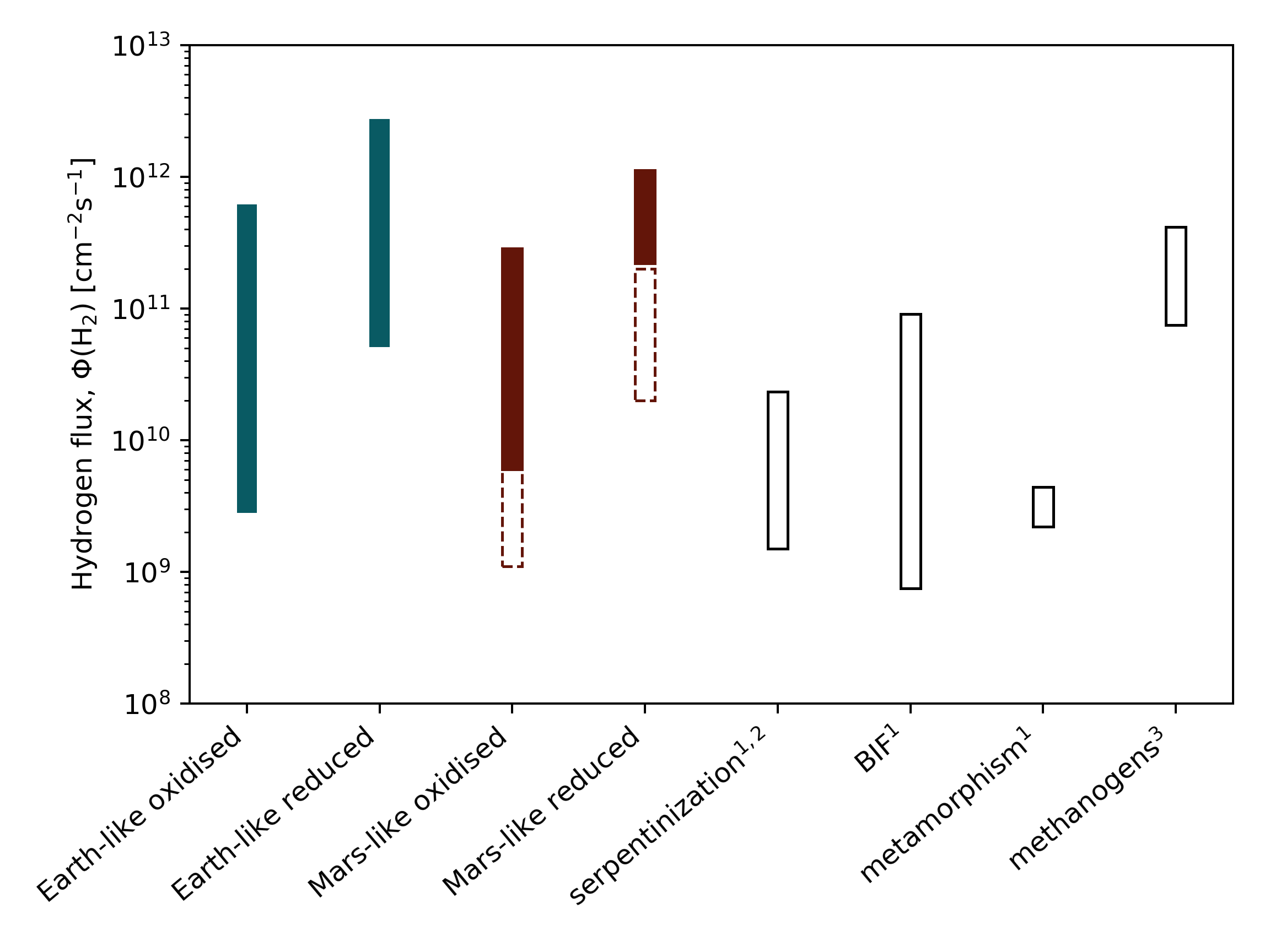}
	\caption{A comparison of the fluxes of alternative sources of \ce{H2} on the early Earth, and volcanic \ce{H2} sources for planets similar to early Earth or early Mars.
	Solid bars represent a volcanic input, the range in possible \fth{} given by our volcanic \ce{H2} model for 4 different conditions.
	Oxidised means FMQ to FMQ-2, reduced is FMQ\,-2.5 to FMQ\,-4.5 (IW\,-1 to IW\,+1).
	Earth conditions are for a range of 1-10\,bar pressure and 1-10$\times$ modern flux using the previously described volatile content.
	Mars data uses the melt composition described in the main text, for vent pressures of 0.5\,-\,2\,bar and 0.5 to 2$\times$ modern Earth volcanic flux.
	For Mars, solid bars represent hydrous conditions, dashed bars the dry conditions.
	Alternative sources of \ce{H2} are represented by open bars. Data taken from: 1. \cite{kasting2013WhatCausedRise}, 2. \cite{batalha2015TestingEarlyMars}, 3. \cite{kharecha2005CoupledAtmosphereEcosystem}.}
	\label{fig:sources}
	\end{center}
\end{figure}

Our calculations so far have assumed that volcanic emissions are the only source of \ce{H2} to the atmosphere in the Archean.
However, while it may be true that the volcanic flux is likely to be the largest steady \ce{H2} input, there are several other sources which may have contributed to the global \ce{H2} budget, summarised in Fig. \ref{fig:sources}.

A particularly large source of uncertainty comes from the sources and sinks of hydrogen introduced by adding life into the model considerations.
\cite{kharecha2005CoupledAtmosphereEcosystem} and \cite{haqq-misra2008RevisedHazyMethane} have suggested that up to 1000\,ppmv \ce{CH4} could be sustained by methanogenic biota in the late Archean.
The \ce{H2} flux this methane production is equivalent to, makes methanogens a potentially equal contributor to volcanism (at high volcanic fluxes) in sustaining atmospheric \ce{H2}.
Methanogens could therefore potentially significantly increase both the total input of hydrogen to the atmosphere and therefore the \fth{}.
However, adding methanogens also produces a sink of hydrogen in the form of organic carbon burial, potentially between $6\times10^{9}$ to $1.5\times10^{10}$\,molecules\,\percmsqpers{} \citep{kasting2013WhatCausedRise}.
Although this is a fairly low flux compared to \ce{H2} sources, it could become significant in cases where the volcanic hydrogen flux is low.
Despite the important supply of \ce{H2} life can provide to an atmosphere, methanogens of course cannot contribute to either a greenhouse atmosphere in a prebiotic system, nor to generating a reducing atmosphere required for prebiotic chemistry. If \ce{H2} atmospheres are needed at this early stage of Earth's history, other sources must be considered.

Other geological sources of hydrogen include the serpentinization of mafic and ultramafic oceanic crust, metamorphic outgassing, and release of \ce{H2} during the formation of banded iron deposits \citep{kasting2013WhatCausedRise}.
Not included in Fig. \ref{fig:sources} is the flux of hydrogen which could be delivered by a large impactor. This has been excluded as an impactor would provide a point source of hydrogen decaying over time, rather than a long-term steady flux, making this a significantly different proposition in comparison to other long term \ce{H2} fluxes.
However, the scale of the flux, a single point injection of possibly $3\times10^{18}$ mols \citep[assuming injection over the course of a year, this would be a point flux of $~ 1\times10^{16}$\,molecules\,\percmsqpers{} for a year;][]{zahnle2019StrangeMessengerNew}, means that impactors could be important for transient high levels of hydrogen in the atmosphere.

\section{Atmospheric Warming of Early Mars}

At 3.8-3.6\,Ga (during the Noachian period), Mars' climate appears to have been warm enough to sustain liquid water at the surface \citep[e.g.,][]{hoke2011FormationTimescalesLarge}.
However, a pure \ce{CO2} Martian atmosphere provides insufficient warming under a faint young sun \citep{forget20133DModellingEarly}.
To address this, \cite{hayworth2020WarmingEarlyMarsa} suggest that a range of \ce{H2} contents of 5 - 8\%, in 1.2 to 3\,bar \ce{CO2} atmospheres, could be sufficient to deglaciate the Martian surface in a cyclical manner.
We therefore perform a similar parameter sweep to that conducted above for Earth, adjusted to be suitable for an early Mars-like planet, to assess the plausibility of this occurring given volcanic constraints.
Here we compare 8\,\% \fth{} in a 1.25\,bar atmosphere to the results of a parameter sweep similar to that conducted above.
We account for the different scale height and diffusion rate of the Martian atmosphere by using a $b_{H_n}/H_a$ ratio taken from \citet{ramirez2014WarmingEarlyMars}, of $1.6\times10^{13}$\,cm$^{-2}\,$s$^{-1}$, which applies for a homopause temperature of $\sim$160\,K.
Using values relevant to 3.8-3.6\,Ga and accounting for the decay in XUV flux with orbital distance \citep[irrandience at Mars is 36-53\% weaker than at 1 AU][]{thiemann2017MAVENEUVMModel}, the incident XUV flux at Mars would have been be in the range $\sim$2.2\,-\,9$\times$ that of the modern Earth, giving a similar range in $\alpha$ to those used for Earth \citep[see Fig.\,\ref{fig:xuv};][]{tu2015ExtremeUltravioletXray}.
We therefore adopt the same range of alpha as used above, accepting that the XUV flux might have a significantly different effect on the escape rate of Mars compared to Earth, with the escape rate possibly being much higher (and therefore a larger value of alpha closer to 1) after scaling for Mars' lower gravity.

\begin{figure}[!ht]
	\begin{minipage}[c]{0.5\textwidth}
	  \includegraphics[width=\textwidth]{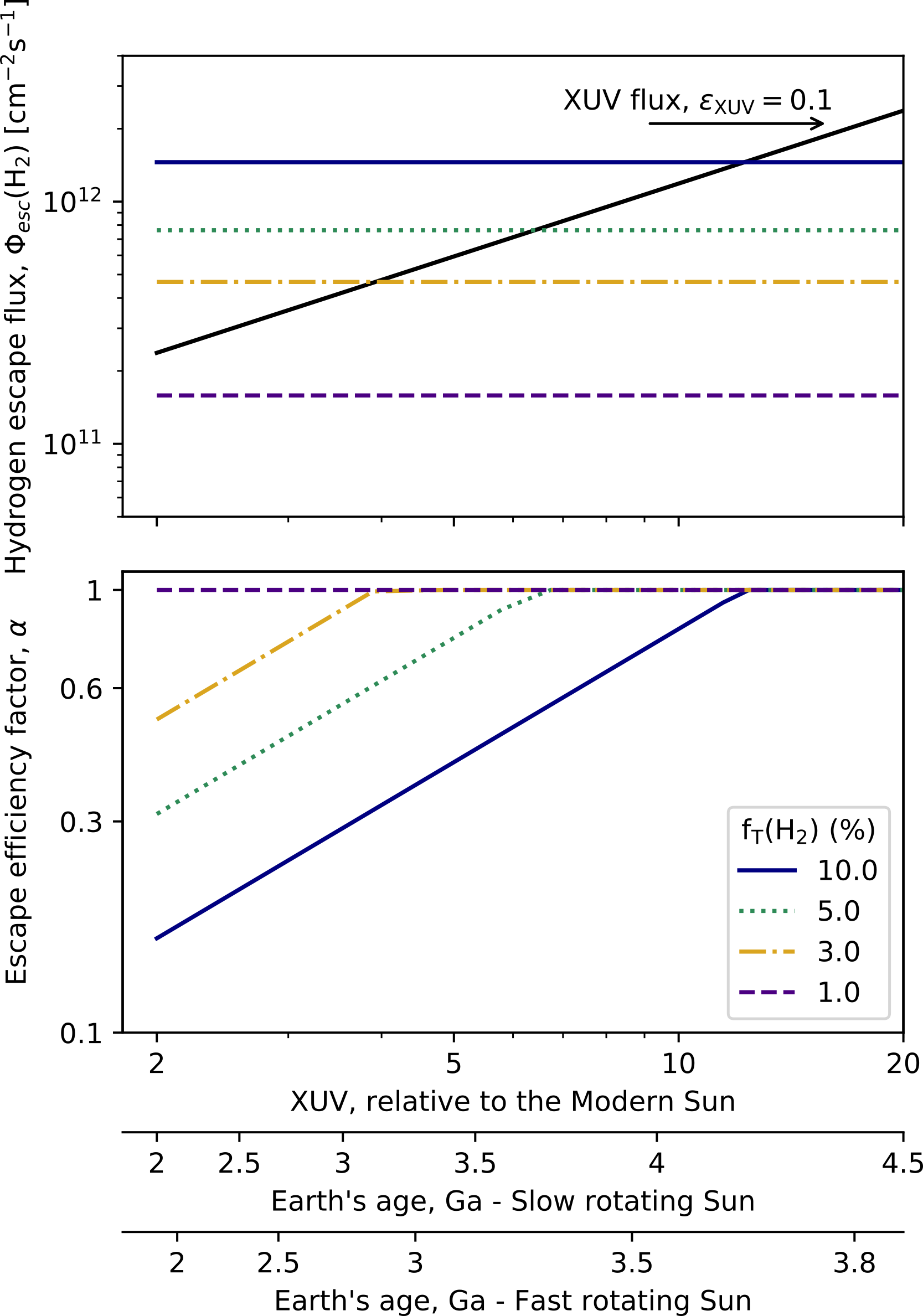}
	\end{minipage}\hfill
	\begin{minipage}[c]{0.4\textwidth}
	  \caption{
		 Top: Comparing the diffusion limited escape rate (horizontal lines) to the energy limited escape rate (diagonal line) calculated using eq.\,\ref{eqn:xuv_flux} Bottom: $\alpha$ calculated as XUV flux/diffusion limited flux.
	  } \label{fig:mars_xuv}
	\end{minipage}
  \end{figure}

We account for the different scale height and diffusion rate of the Martian atmosphere by using a $b_{H_n}/H_a$ ratio taken from \citet{ramirez2014WarmingEarlyMars}, of $1.6\times10^{13}$\,cm$^{-2}\,$s$^{-1}$, which applies for a homopause temperature of $\sim$160\,K.
Although the XUV flux at Mars is 36-53\% weaker than at 1 AU \citep{thiemann2017MAVENEUVMModel}, at 3.6-3.8\,Ga the solar XUV flux is approximately double that of the period modelled for Earth.
Both of these factors conspire to decrease the difference between the diffusion and energy limited flux regimes on Mars at the relevant time period, leading to increased $\alpha$.
By comparing the diffusion limited flux with XUV-limited escape using Eq.\ref{eqn:xuv_flux} and $\epsilon_{\rm XUV} = 0.1$, we plot $\alpha$ in Fig \ref{fig:mars_xuv}.

\begin{figure}[H]
	\begin{center}
	\captionsetup[subfigure]{justification=centering}
	\begin{subfigure}{.6\textwidth}
		\centering
		\includegraphics[width=\linewidth]{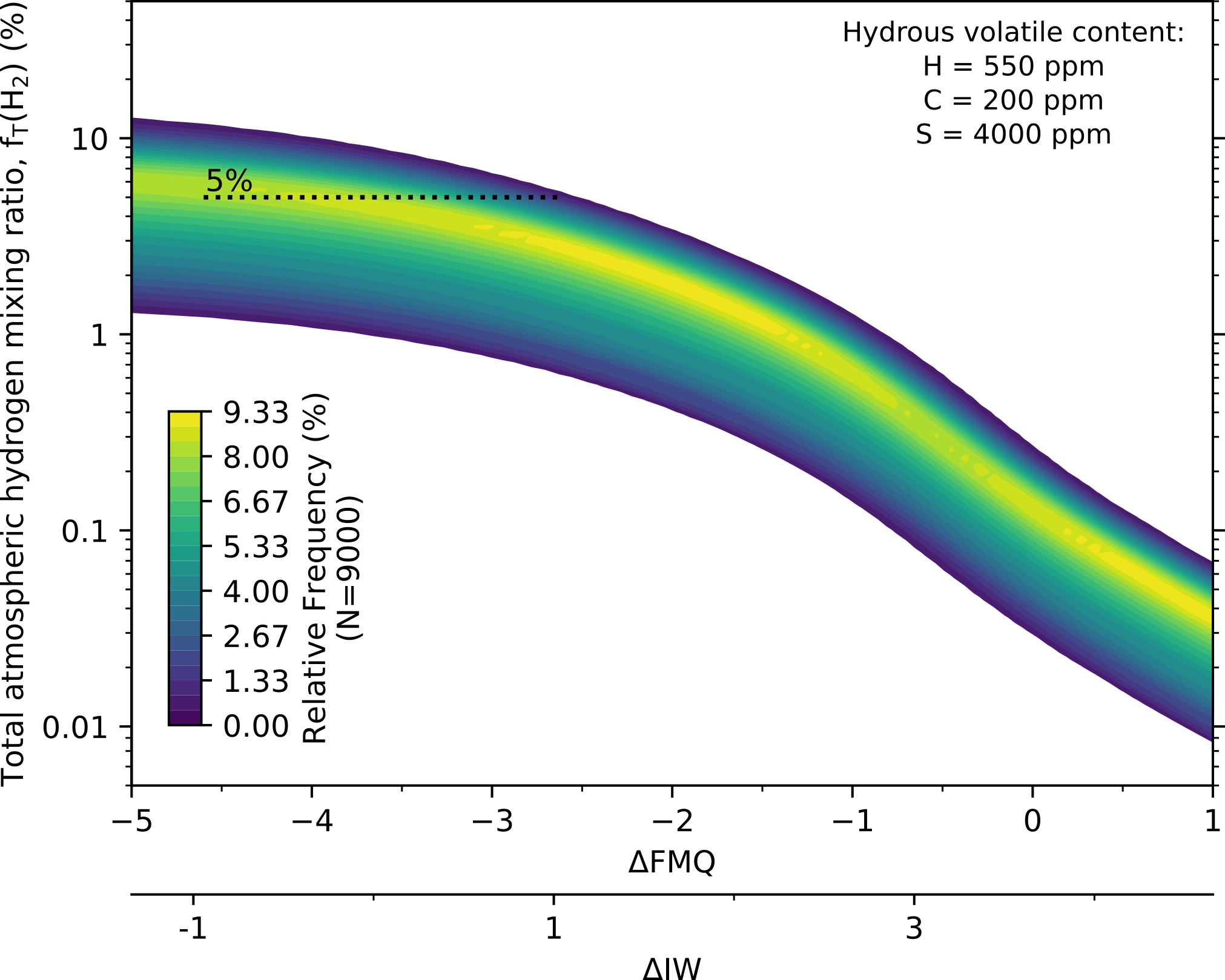}
		\caption{}
		\label{fig:prob_m_cont}
	\end{subfigure}\hfill
	\begin{subfigure}{.3\textwidth}
		\centering
		\includegraphics[width=\linewidth]{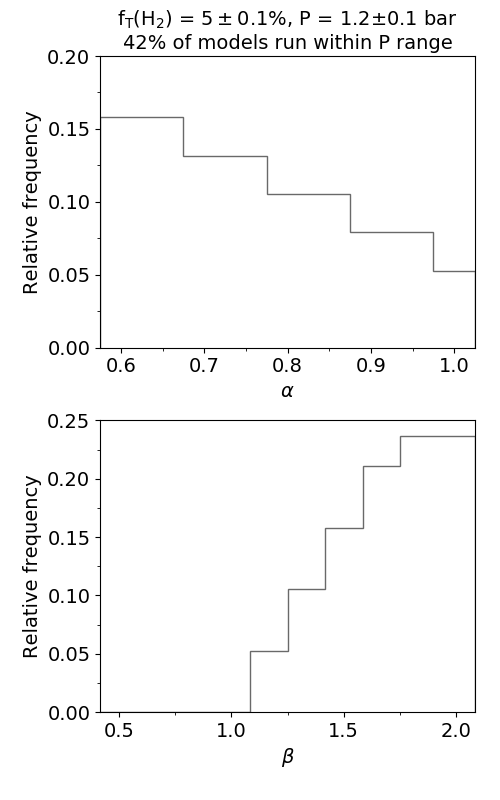}
		\caption{}
		\label{fig:m_priors}
	\end{subfigure}%
	
	\begin{subfigure}{\textwidth}
		\includegraphics[width=.6\textwidth]{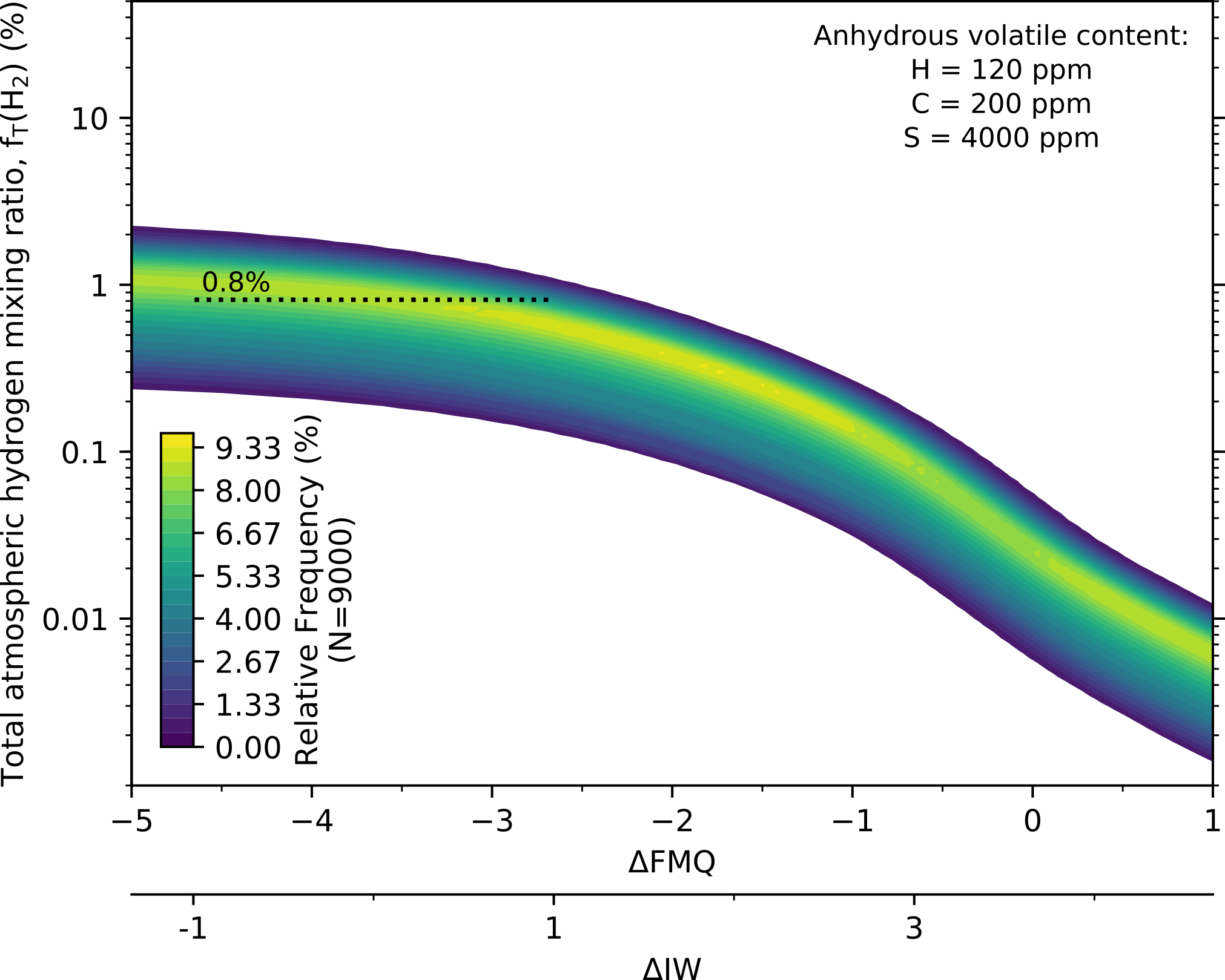}
		\caption{}
		\label{fig:prob_m_anhy}
	\end{subfigure}\hfill


	\caption{(a) The relative frequency of model predictions of \ce{H2} mixing ratios in the atmosphere of a Mars-like planet under the hydrous scenario as a function of the magma \fo{}, represented as a 2D density surface, assuming: $\alpha = 0.6-1$, $\beta$ of 0.5 - 2 and a surface pressure of 0.5 - 2\,bar \citep{kurokawa2018LowerLimitAtmospheric,warren2019ThickThinNew}.
	White regions are unphysical.
	Horizontal bar indicates the estimated Martian \fo{} range.
	(b) Relative frequency distributions for the two parameters varied in the calculations presented in panel (a), for models which can achieve 5\,\% \fth{} at 1.2 $\pm$ 0.1\,bar pressure within the \fo{} range of Mars, labelled with the percentage of model runs which meet these constraints.
	(c) Relative frequency plot equivalent to (a) but for a dry scenario.}
	\label{fig:mars_prob}
\end{center}	
\end{figure}

If we assume that the \ce{H2} loss rate proceeds exactly at the XUV limited rate until it intersects the diffusion limit (Fig.\,\ref{fig:mars_xuv}), for the period 3.8-3.6\,Ga $\alpha$ would only be $<$\,1 when \fth{} $\geq\,5$\,\%, with the greatest reduction below the diffusion limit being $\alpha$=0.6 for \fth{}\,=\,10\,\%. However, this is not quite accurate; the true \ce{H2} loss rate is likely to drop below the XUV limit as the flux asymptotes towards the diffusion limit, prolonging the period where $\alpha$\,=\,0.7\,-\,0.9 beyond that seen in Fig.\,\ref{fig:mars_xuv} (see Fig\,\ref{fig:xuv}). We therefore use a range of 0.6\,-\,1 for $\alpha$ in Fig.\,\ref{fig:mars_prob}.

The volcanic fluxes considered are 0.5-2$\times$ the modern Earth, in keeping with the idea that the heat flux in the Noachian was probably similar to the modern Earth's \citep{ramirez2014WarmingEarlyMars,batalha2015TestingEarlyMars}.
On this basis, we consider that the volume of magma reaching the surface per unit area should therefore be similar.
However, this assumption relies on similar magmatic volatile contents between the two planets, and the primary volatile contents of early Martian magmas are very poorly constrained.
Estimates of the pre-eruptive water contents of Martian magmas have ranged from nearly anhydrous through to around 2\,wt\,\% based on analysis of the SNC meteorites \citep[e.g.,][]{mcsween2001GeochemicalEvidenceMagmatic, usui2012OriginWaterMantle, mccubbin2012HydrousMeltingMartian,filiberto2016ConstraintsWaterChlorine}, which sample magmas which were erupted at $<$\,2\,Ga \citep{nyquist2009ConcordantRbSr}.
While differences in analytical technique may explain some of this variation, there are a large number of unknowns for Mars which could contribute to this uncertainty, including the extent of volatile recycling \citep[e.g.,][]{magna2015CalciumIsotopeSystematics, batalha2015TestingEarlyMars}, heterogeneous distributions of water in the mantle \citep{mccubbin2016HeterogeneousDistributionH2O}, the potential for volatile enrichment in magmas via crustal assimilation \citep{mccubbin2016HeterogeneousDistributionH2O}, and the evolution of mantle (and therefore magmatic) water contents through time \citep[e.g.,][]{filiberto2016ReviewVolatilesMartian}.

In light of the large uncertainty around the water contents of Martian magmas, we ran two parameter sweeps to find upper and lower limits on the potential atmospheric hydrogen content of Noachian Mars.
The hydrous scenario, valid if volatile cycling (via crustal assimilation or other processes) or a large initial mantle water budget was present, is run with 550\,ppm H, 200\,ppm C, and 4000\,ppm S, equivalent to 0.3\,wt\,\% \ce{H2O} and 600\,ppm \ce{CO2} dissolved prior to degassing at IW \citep[e.g., upper limit of ][]{mccubbin2012HydrousMeltingMartian}.
The dry scenario is run with equal C and S contents, and with 120\,ppm H, equivalent to 0.03\,wt\,\% \ce{H2O} and 400\,ppm \ce{CO2} \citep{mccubbin2016HeterogeneousDistributionH2O, filiberto2016ConstraintsWaterChlorine}.
While the C content of Mars is equally uncertain, a variation in the magma carbon content of several hundred ppm would have a negligible effect on \fth{} at $\sim$ 1\,bar pressure (Fig.\,\ref{fig:mixing ratio}) so the C content between the two runs remains fixed.
Our results are summarised in Fig. \ref{fig:mars_prob}.

Fig.\,\ref{fig:mars_prob} indicates that across the likely range of magmatic \fo{} on Mars, the range of \ce{H2} mixing ratios predicted for the hydrous scenario is $\sim$0.6 - 11\,\% \ce{H2}, with most models falling in the range 2 to 8\,\% \fth{} (0.9 - 15\,\% accounting for photochemical uncertainty, see \ref{appendix_param}). We find that 42\,\% of models run at 1.2 $\pm$ 0.1\,bar achieve a 5\,\% \fth{} within Mars' likely magmatic \fo{} range.
The number of hydrous model runs which achieve 5\,\% \fth{} decreases as $\alpha$ increases, and Fig.\,\ref{fig:mars_prob}\,(b) shows a strong bias towards high values of $\beta$; no model runs within Mars' \fo{} range meet \fth{}\,=\,5\,\% with $\beta\leq$\,1.
Under a dry scenario, the most frequent results fall to 0.4 - 1.2\,\% \fth{} (0.1 - 3 with photochemical uncertainty, see \ref{appendix_param}).
Given the relative frequency with which hydrous degassing models with Martian parameters can generate \fth{}\,=\,5\,\%, it is plausible that a volcanically-driven hydrogen greenhouse effect contributed to the warm and wet Noachian period on Mars, particularly if erupting magmas are at or below the IW buffer (Fig.\,\ref{fig:mars_prob}\,(a)).
While under dry conditions this seems to be highly unlikely, as with Earth, there may also have been other geological sources of \ce{H2} to the Martian atmosphere, such as weathering of white rust \citep[\ce{Fe(OH)2}, ][]{tosca2018MagnetiteAuthigenesisWarming}.

A key requirement of \ce{H2}-driven warming on Mars is that the atmosphere is dominantly \ce{CO2}. However, under reducing conditions with a high S content and a hydrous melt, most of the gas emission occurs as \ce{H2} and \ce{H2S}, with C-bearing species only making up $\sim$5\,\% of the gas emitted (as a molar fraction).
\cite{ramirez2014WarmingEarlyMars} suggest that the \ce{CO2} in the atmosphere could be sourced from volcanism through indirect \ce{CO} and \ce{CH4} reactions, but if less than 5\,\% of the volcanic emissions are carbon-bearing species, then it may pose a challenge for a \ce{CO2}-\ce{H2} atmosphere to be maintained solely through volcanism.

\cite{ramirez2014WarmingEarlyMars} also suggest that the atmospheric \ce{CO2} could have been sourced earlier in Mars' history through the initial solidification of the upper 50\,km of Mars' crust.
However, carbon in Mars' atmosphere is thought to have been lost via hydrodynamic escape prior to 4.1\,Ga, suggesting a \ce{CO2}-rich atmosphere from crustal solidification could not have been maintained \citep{tian2009ThermalEscapeCarbon}.
One possible solution to this is an intrinsically more carbon-rich martian mantle;  this would have a minimal effect on the hydrogen fraction produced (Fig.\,\ref{fig:mixing ratio}), but could increase the flux of C-bearing species to the atmosphere.
As the constraints on the volatile content of the Martian interior are very uncertain, it is difficult to rule this out.

\section{Conclusions}

We have investigated the possibility of volcanism sustaining secondary atmospheres with a high \ce{H2} mixing ratio.
We have found that \fth{} values of several percent can be maintained, even with relatively oxidised magmatic sources such as those seen on Earth.
As the magmatic \fo{} decreases, the \fth{} sustained increases, to the point where the assumption that \ce{H2} is a minor species breaks down as \fth{} approaches 50\,\%.
These high \ce{H2} fractions are very unlikely to be formed where the volcanic degassing occurs under high atmospheric pressure or underwater, and for more oxidising magmas a greater volcanic flux and/or a less efficient \ce{H2} loss to space is necessary to maintain significant atmospheric \ce{H2} fractions.

We have found that a sustained $\sim$1\,\% hydrogen atmosphere is plausible for the early Earth, given the likely range of magmatic and surface conditions, and does not require a mantle source reduced beyond that seen today.
However, \ce{H2} mixing ratios at the 10\% level cannot be achieved given the likely conditions and magmatic \fo{} on the early Earth, without supply from another source or suppression of escape beyond that which we consider is plausible. \ce{H2}-enhanced warming is therefore unlikely to have significantly contributed to counteracting the faint young sun.
In the case of Mars, we suggest that sufficient \ce{H2} could be degassed to carry out greenhouse warming if primary Martian magmas are relatively water-rich. However, the challenge under this scenario is that there may not have been a sufficient \ce{CO2} source for this greenhouse mechanism to be viable; only $\sim$\,5\,mol\,\% of volcanically degassed species are carbon-bearing under hydrous conditions according to our model, and any \ce{CO2} from early crustal solidification is likely to have already been lost. The sensitivity of these results to primary magmatic water contents shows the importance of improving constraints on the volatile content of Mars' mantle, for understanding the planet's early climate.

\section{Acknowledgements}

With thanks to an anonymous reviewer and Fabrice Gaillard for helpful comments. The authors would like to acknowledge C.E ``Sonny" Harman and Mark Claire for helpful conversations. P.\,B.\,R thanks the Simons Foundation for support under SCOL awards 59963. Funding: This work was supported by the Embiricos Trust Scholarship from Jesus College, Cambridge.

\bibliographystyle{my_epsl} 
\bibliography{bibl}

\newpage

\appendix
\section{Effect of pressure and \fo{} on volcanic gases and \ftvh{}} \label{appendix_outgass}

\subfile{appendixA}
\section{Outgassing parameterisation} \label{appendix_param}

\subfile{appendixB}

\end{document}

%% file: appendixA.tex
\begin{figure}[H]
	\begin{center}
	\includegraphics[width=0.6\textwidth]{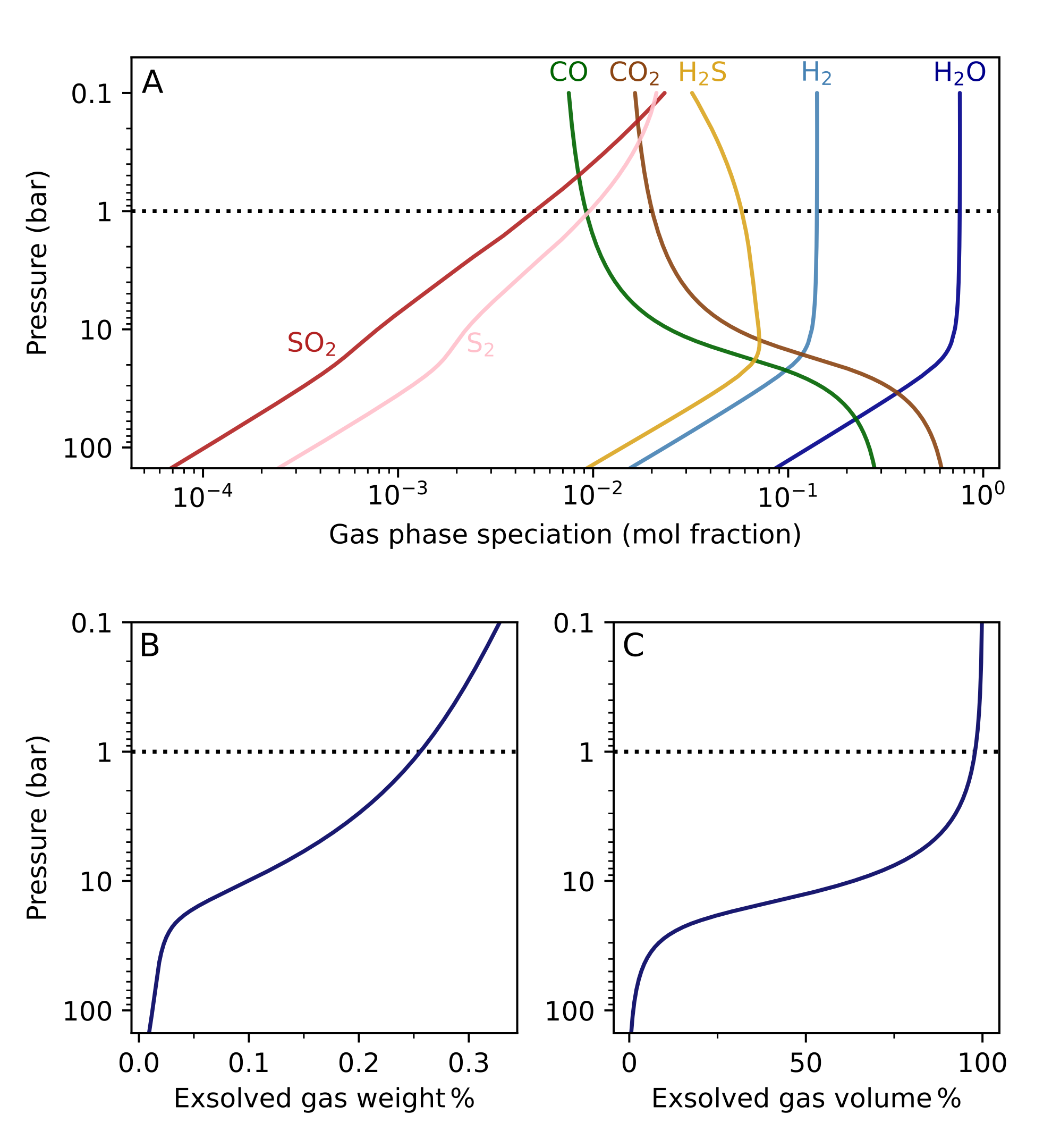}
	\caption{A: Speciation of the gas phase from a melt at FMQ\,-2, with an initial volatile content of 350\,ppm \ce{H}, 50\,ppm C and 1000\,ppm S, at a range of final pressures. B: Weight\,\% of the same system made up of the gas phase according to pressure. C: Volume\,\% of the system taken up by the gas phase according to pressure. Dotted lines correspond to the same transect on Figure \ref{fig:hor_spec_350}.}
	\label{fig:depth_spec}
	\end{center}	
\end{figure}
\begin{figure}[H]
	\begin{center}
	\captionsetup[subfigure]{justification=centering}
	\begin{subfigure}{.5\textwidth}
	  \centering
	  \caption{350\,ppm H, 1\,bar}
	  \includegraphics[width=.9\linewidth]{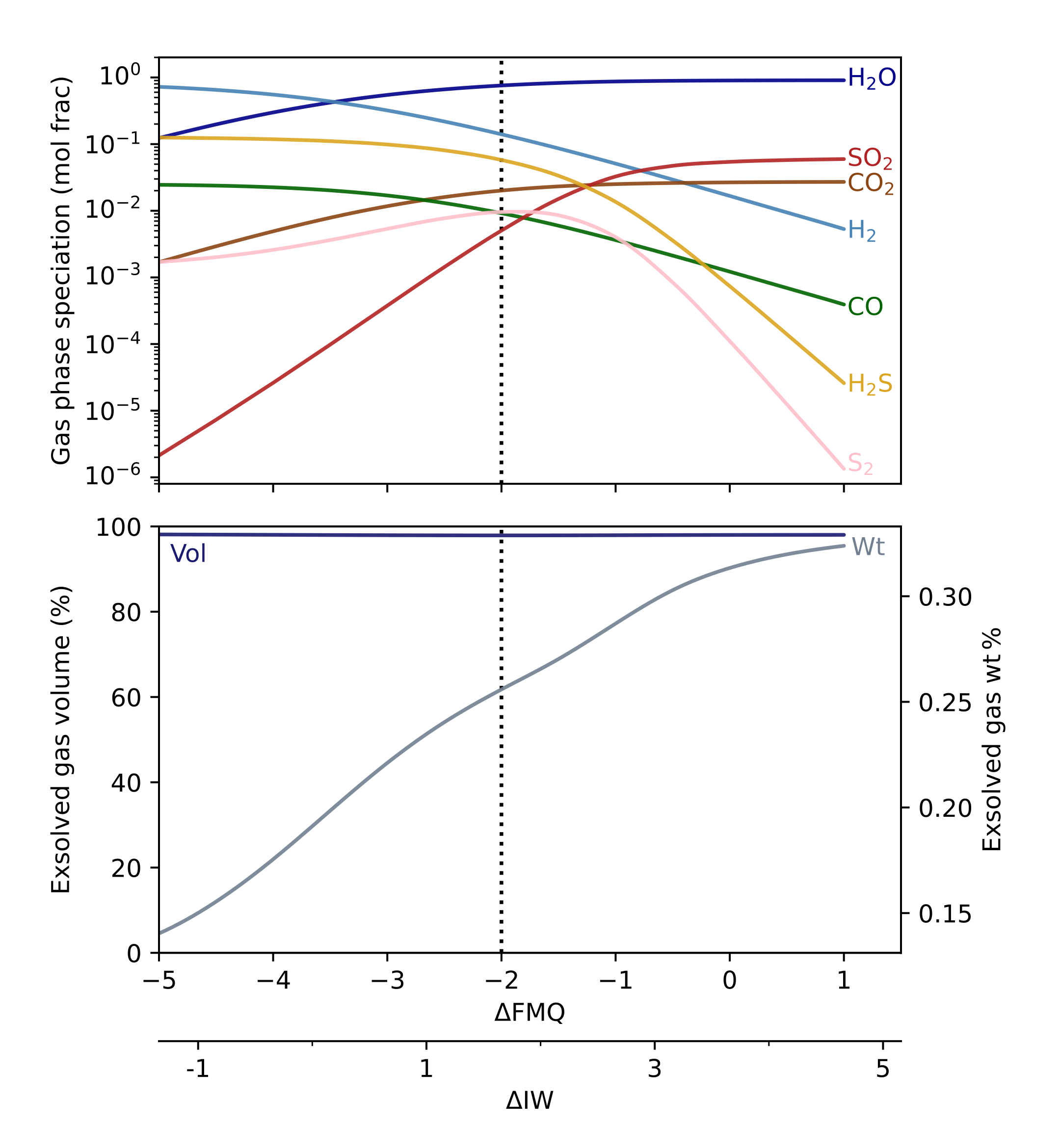}
	  \label{fig:hor_spec_350}
	\end{subfigure}%
	\begin{subfigure}{.5\textwidth}
	  \centering
	  \caption{800\,ppm H, 100\,bar}
	  \includegraphics[width=.9\linewidth]{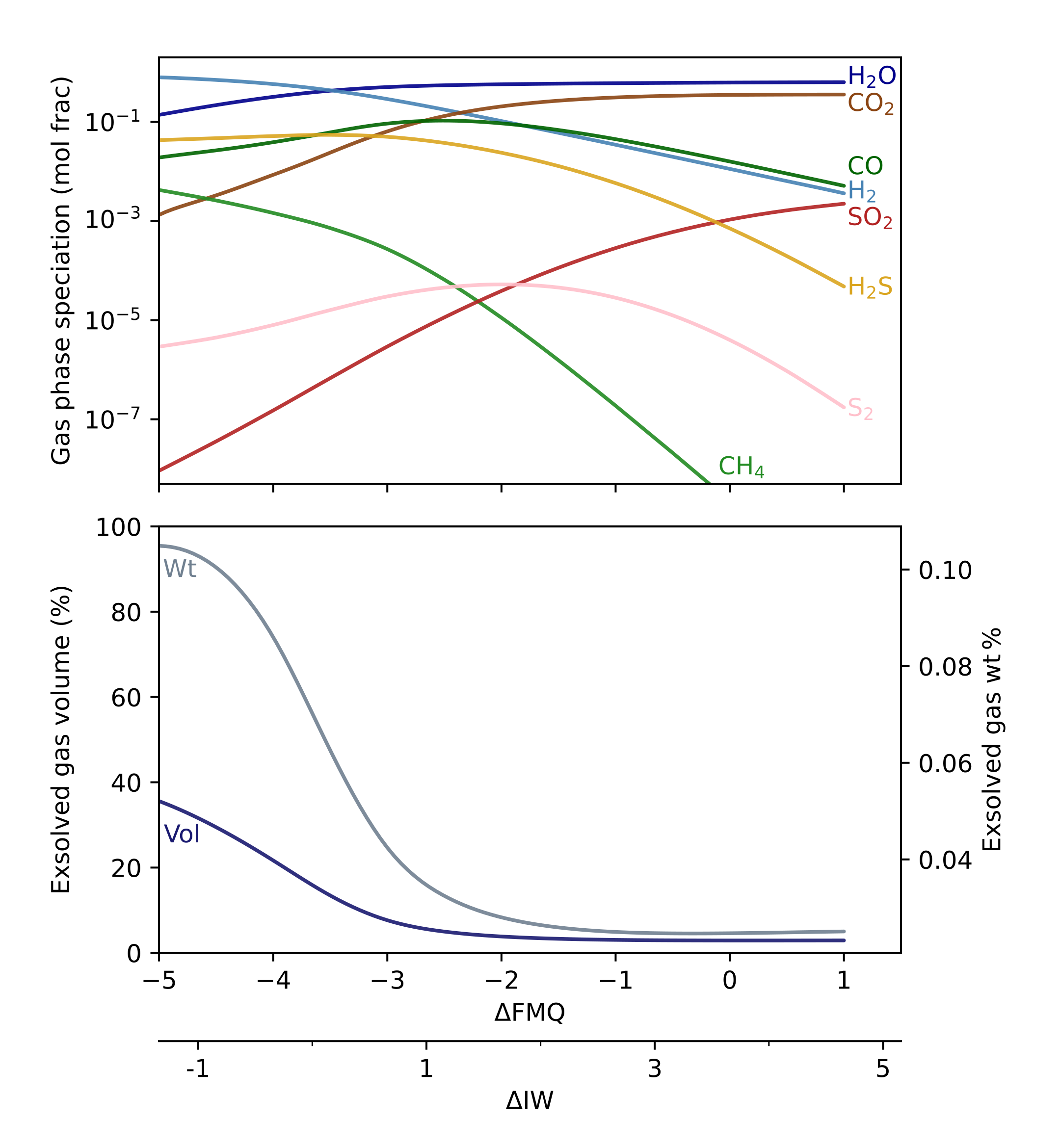}
	  \label{fig:hor_spec_800}
	\end{subfigure}%
	\caption{Top: Speciation of the gas phase at 1 bar from a melt at a given \fo{}, with an initial volatile content of 50\,ppm C, 1000\,ppm S and a specified H content.
	Bottom: Changing gas weight percent and volume percentage of the system taken up by the gas phase. Dotted lines match a corresponding transect point on Figure \ref{fig:depth_spec}.}
	\label{fig:hor_spec}
\end{center}	
\end{figure}

The amount of gas released from a melt and it's speciation varies with pressure (Figure \ref{fig:depth_spec}(a, b)). As the pressure reduces, so does volatile solubility therefore increasing the volume of the gas phase. The volume of gas phase also increases according to the ideal gas law, so that Fig.\ref{fig:depth_spec}c shows both the increasing amount of volatile in the gas phase, and the expansion of the gas phase according to $PV=nRT$ as the pressure decreases. Figure \ref{fig:depth_spec}a shows how the speciation of a gas phase changes with depth, but it should be noted that this is for a fixed \fo{} at all pressures, to isolate the pressure effect on speciation. If left to decompress naturally, a melt (particularly a sulphur rich one) will evolve through a slight \fo{} change and the resulting speciation plot would be a combination of effect from Figures \ref{fig:depth_spec}a and \ref{fig:hor_spec}.

The exsolved fraction also varies with constant pressure and variable \fo{} (Fig.\ref{fig:hor_spec}). As the \fo{} decreases at low pressure, the weight fraction of gas exsolved decreases, while the gas volume remains almost constant. The decreasing weight fraction reflects oxygen in the gas phase reducing with \fo{}. As oxygen is the second heaviest volatile element, reducing the system \fo{} results in a gas phase which is dominated by lighter molecules and therefore has a much lower mean molecular mass. The decreasing weight fraction from oxygen loss is modulated by the presence of sulphur, which initially partitions slightly more into the melt, before increasing in the gas phase as it speciates into less soluble, reducing species (\ce{S2} and \ce{H2S}) rather than \ce{SO2}.

However at high pressures (Fig.\,\ref{fig:hor_spec_800}), the amount of volatile in the gas phase clearly increases from very low fractions with decreasing \fo{} due to the solubility of a species being linked to it’s fugacity in the gas phase (Eq.\ref{eqn:sol law}).
As the \fo{} decreases and water in the gas phase dissociates into \ce{H2} (or is taken up as \ce{H2S} in H-poor systems), more water is drawn out of the melt to maintain the \ce{H2O$_{melt}$ \rightleftharpoons H2O$_{gas}$} equilibrium.
The resulting hydrogen does not dissolve back into the melt, as \ce{H2} and \ce{H2S} have a much lower solubility than water.
Similarly, \ce{CO2} is converted to insoluble \ce{CO} and \ce{CH4} in the gas phase.
The increase in the volatile fraction with decreasing \fo{} is less obvious at low pressures (Fig.\,\ref{fig:hor_spec_350}), as almost all the volatile elements are already in the gas phase, so the absolute change in volume with \fo{} is much smaller.
With a H-rich magma such as in Fig.\,\ref{fig:hor_spec_800}, the lower mean molecular mass of the gas phase is insignificant compared to the amount of volatile exsolving from the melt, so the trend for the weight fraction of gas to decrease is overridden.

There are therefore two factors (pressure and \fo{}) which can modulate the amount of volatile released from a magmatic system, independent of the initial volatile load.
These two factors interact, so for any melt with a fixed H, C and S content and the same rate of magma flux (mantle heat flow) to the surface, it's position in \fo{} and pressure space once it erupts will control the volcanic gas flux (in mol\,s$^{-1}$).

\begin{figure}[H]
	\begin{center}
	\includegraphics[width=0.9\textwidth]{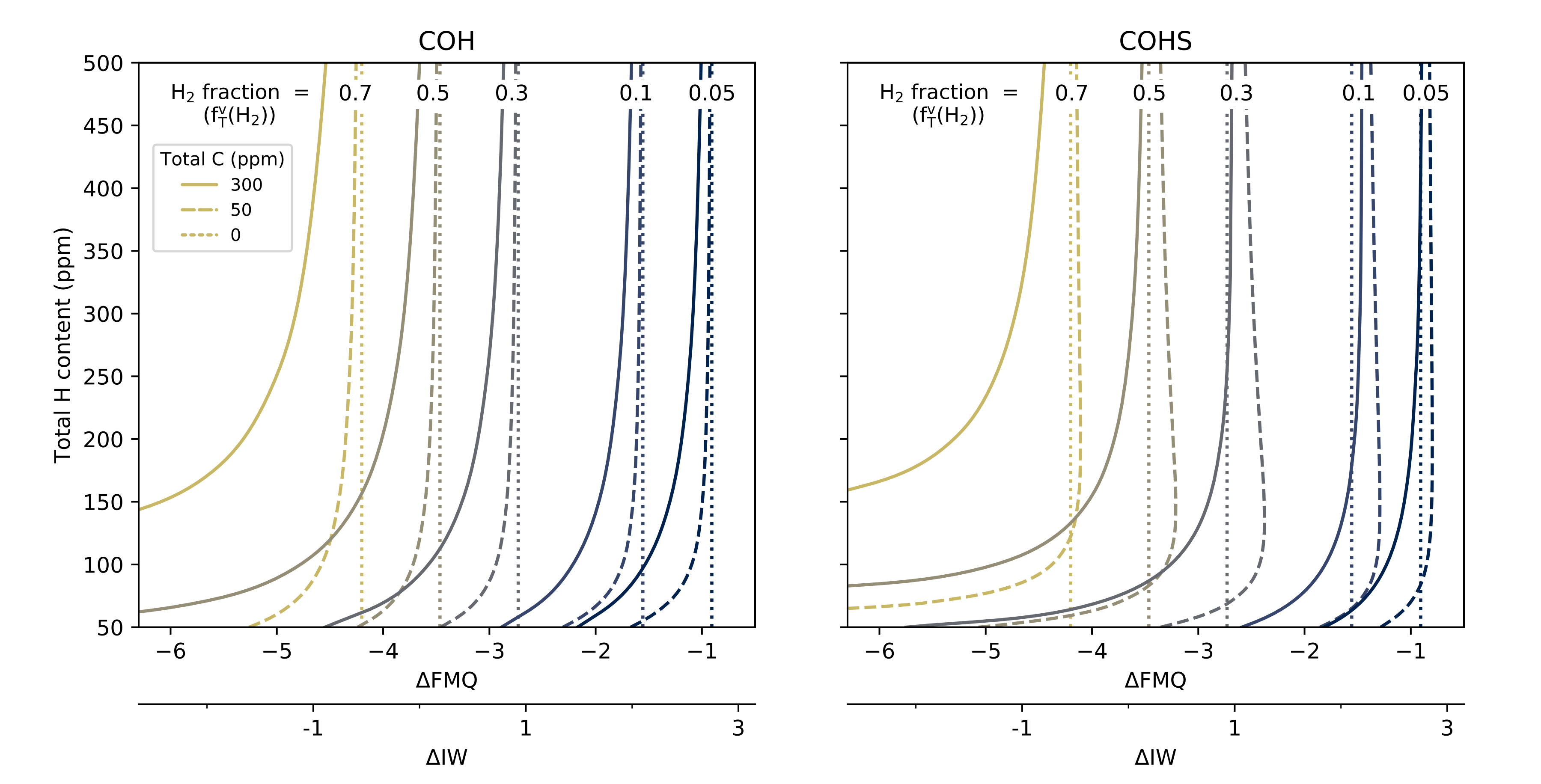}
	\caption{Mole fraction of the gas phase occurring as total \ce{H2} at 1\,bar, according to the \fo{} of the melt at the surface, and the amount atomic hydrogen and carbon dissolved in the melt at depth.}
	\label{fig:h2_contour}
	\end{center}
\end{figure}

The mixing fraction of total hydrogen in a volcanic gas (\ftvh) emitted at 1\,bar surface pressure can be seen to vary with the melts initial volatile content, along with the \fo{} of the erupting melt with reference to the FMQ and IW buffers (Figure \ref{fig:h2_contour}).

As the total carbon content of the melt increases, a progressively more reduced system is required to achieve the same \ftvh{}.
In these melts, a greater fraction of the gas volume is made up of carbon bearing species, mostly \ce{CO2} and/or CO, which both have low to no solubility in magma at low pressures.
Therefore, almost all of the volatile carbon in the system is in the gas phase at 1 bar (see Fig.\ref{fig:depth_spec}) and a greater C content means a larger absolute amount of reduced, H-bearing species are required to achieve the same \ftvh{}. 

In a H-rich COH system, the gas phase is predominantly \ce{H2O} (\ce{H2} at very low \fo{}), whereas in H-poor systems with the same \fo{}, \ce{CO2} and \ce{CO} are much more dominant.
Therefore, to achieve the same \ftvh{}, a greater proportion of the water in the system has to be dissociated.
With a large volume of water, this dissociation happens easily, but as the available water fraction decreases, a greater reduction in \fo{} is required to drive the dissociation in eq.\ref{eqn:k1}, resulting in the curve towards a lower \fo{} as the H fraction decreases.
A reduced H content also means there is a lower volatile fraction in the system as a whole, and therefore the C-bearing species make up a greater proportion of the total gas phase.
In H-poor magmas, the \ftvh{} contours asymptote to the fraction of the gas phase made up of H-free, carbon bearing species; the gas phase is then almost entirely composed of \ce{H2}, \ce{CO} and at higher \fo{}, \ce{CO2}.
With higher C contents, this H-free fraction is larger and so the maximum achievable \ftvh{} occurs at a correspondingly greater H content.

Adding sulphur to the system has the effect of shifting the contours to a slightly higher \fo{}, and increasing the H-free gas fraction again to shift the point of maximum achievable \ftvh{} to a higher H content.
Whereas in the COH system the only species in the \ftvh{} fraction are \ce{H2} and \ce{CH4}, which are both present in low to very low fractions at high to moderate \fo{}, in the COHS system the \ftvh{} fraction includes \ce{H2S}.
As \ce{H2S} can be present in significant fractions at all \fo{} values discussed here, adding S to the system makes achieving a specific \ftvh{} easier than in both the COH and OH systems.
The effect of adding S to a system is greatest at around FMQ-3, where the fraction of \ce{H2S} approaches its maximum value before much of the \ce{H2O} has strongly dissociated into \ce{H2} (Fig.\ref{fig:hor_spec}).
This effect is particularly strong in H-poor systems, where \ce{H2S} becomes the dominant \ftvh{} species (as opposed to \ce{H2}) at low \fo{}.

%% file: appendixB.tex
\begin{figure}[H]
	\begin{center}
	\includegraphics[width=0.9\textwidth]{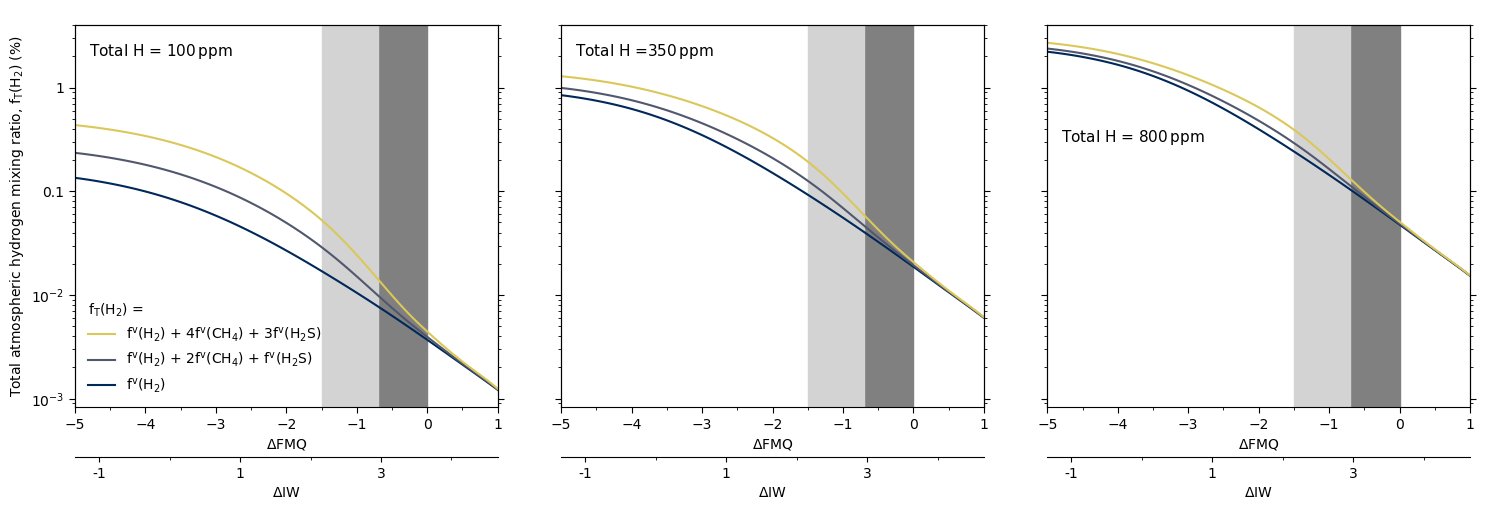}
	\caption{The atmospheric \fth{} variation with \fo{} in a one bar atmosphere, 550\,ppm\,C, according to the volatile content and \ftvh{} stoichiometry. Light and dark grey bars represent the possible Archean range and modern MORB range of \fo{}.}
	\label{fig:stoichiometry}
	\end{center}
\end{figure}

Depending on how the volcanic hydrogen input \ftvh{} is defined, the same volcanic conditions an generate a different atmospheric \fth{} (Fig. \ref{fig:stoichiometry}). 

At the lower bound of \ftvh{}, only hydrogen emitted as molecular \ce{H2} is counted. This ignores any input from other hydrogen bearing species, and assumes they are all removed from the atmosphere via mechanisms which do not release \ce{H2}. This could be a result of either very efficient \ce{H2S} deposition, or a sufficiently oxidising atmosphere such that 

\begin{equation}
	\ce{H2S} + \frac{3}{2}\ce{O2} \leftrightharpoons \ce{SO2} + \ce{H2O},
\end{equation}
preserving atmospheric redox. However, the Archean atmosphere pre-GOE is very unlikely to have been oxidising enough to make this plausible.

On the other end of the scale, the upper bound for \ftvh{} is where all the \ce{CH4} and \ce{H2S} is included, weighted according to how much hydrogen will be emitted by an oxidation reaction with each molecule.
I.e., one mol of volcanic \ce{CH4} is equal to 4 mols of \ce{H2} in the atmosphere as it reacts according to

\begin{equation}
	\ce{CH4} + 2\ce{H2O} \leftrightharpoons \ce{CO2} + 4\ce{H2},
	\end{equation}
and \ce{H2S} produces 3 mols of \ce{H2} according to
\begin{equation}
	\ce{H2S} + 2\ce{H2O} \leftrightharpoons \ce{SO2} + 3\ce{H2}.
	\end{equation}
This would suggest inefficient \ce{H2S} deposition and a neutral atmosphere so that \ce{H2S} and \ce{CH4} react sufficiently quickly in the troposphere and lower stratosphere such that

\begin{equation}
	f^{\mathrm{v}}_{\mathrm{T}}(\mathrm{H}_2) = f(\ce{H2})+ 4f(\ce{CH4}) + 3f(\ce{H2S}).
	\end{equation}

As seen in Figure \ref{fig:stoichiometry}, the middle ground can be represented by 
\begin{equation}
    f^{\mathrm{v}}_{\mathrm{T}}(\mathrm{H}_2) \approx f^{\rm v}(\ce{H_2}) + 2f^{\rm v}(\ce{CH4)} + f^{\rm v}(\ce{H2S}).
\end{equation}
This is the same stoichiometry as is used to define \fth{} above the tropopause weighting \ce{CH4} and \ce{H2S} according to the number of H atoms they contain, and approximates the idea that a proportion of \ce{CH4} and \ce{H2S} will react to \ce{H2}, with the rest being lost through alternative mechanisms such as an environment where \ce{H2S} and \ce{CH4} deposition may be inefficient, but they are also less reactive in the atmosphere either because the atmosphere is neutral and there is insufficient stellar UV to convert to \ce{SO2}/\ce{CO2} before reaching the homopause.
Alternatively, species are photochemically restored as soon as they are reacted away in a reducing atmosphere.
Changing the stoichiometry of \ftvh{} has the greatest effect at low water contents, where there is a higher ratio of \ce{H2S}:\ce{H2}, where the greatest difference between the upper and lower bound is 3.5$\times$ at 100\,ppm, to 1.3$\times$ at 800\,ppm \ce{H2}.

\begin{figure}[H]
	\begin{center}
	\captionsetup[subfigure]{justification=centering}
	\begin{subfigure}[t]{.3\textwidth}
	  \centering
	  \includegraphics[width=\linewidth]{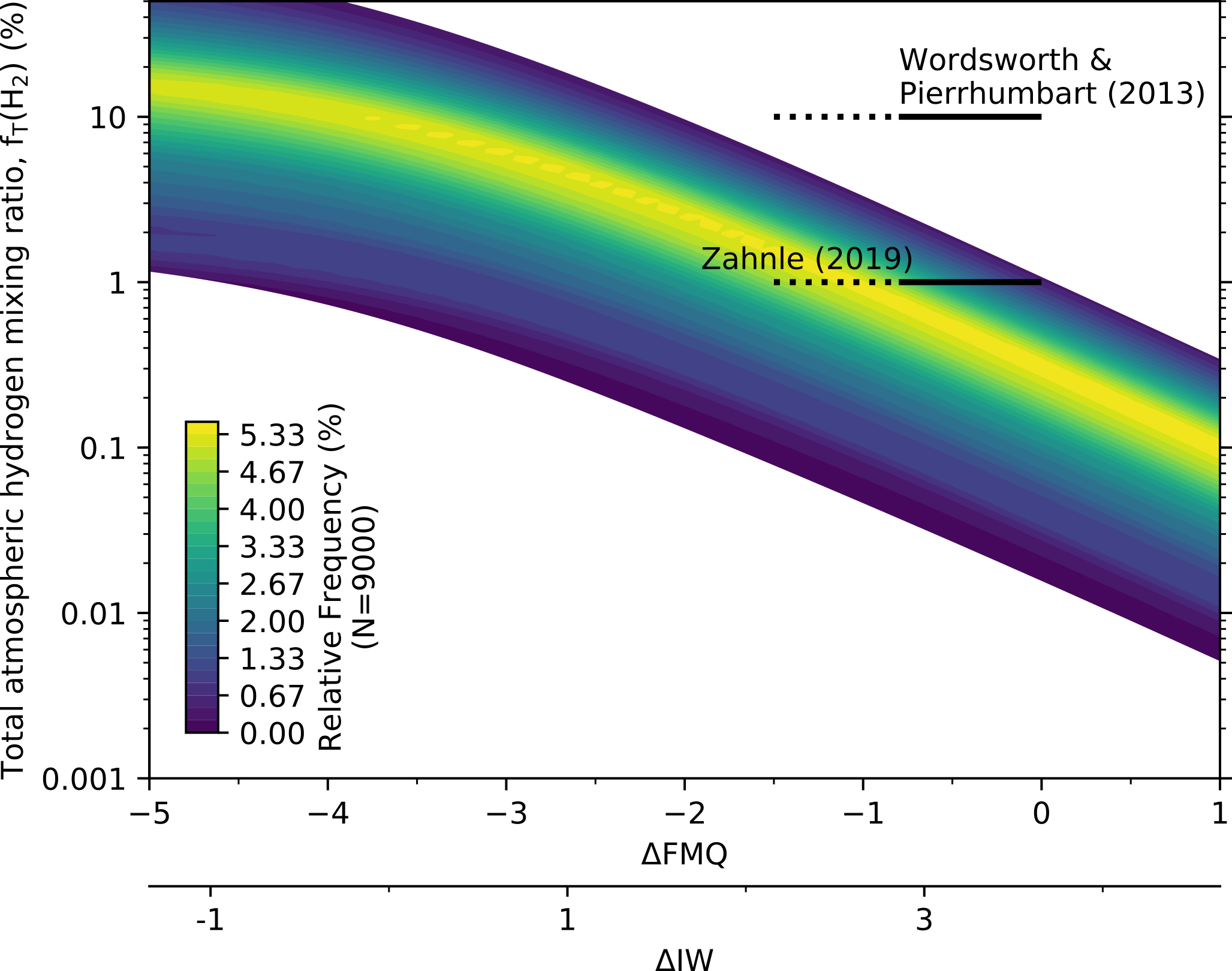}
	  \caption{$f^{\rm v}(\ce{H_2})$} 
	  \label{fig:prob_e_low}
	\end{subfigure}\hfill%
	\begin{subfigure}[t]{.3\textwidth}
		\centering
		\includegraphics[width=\linewidth]{earth_norm_H.png}
		\caption{$f^{\rm v}(\ce{H_2}) + 2f^{\rm v}(\ce{CH4)} + f^{\rm v}(\ce{H2S})$} 
		\label{fig:prob_e_norm}
	  \end{subfigure}\hfill%
	\begin{subfigure}[t]{.3\textwidth}
	  \centering
	  \includegraphics[width=\linewidth]{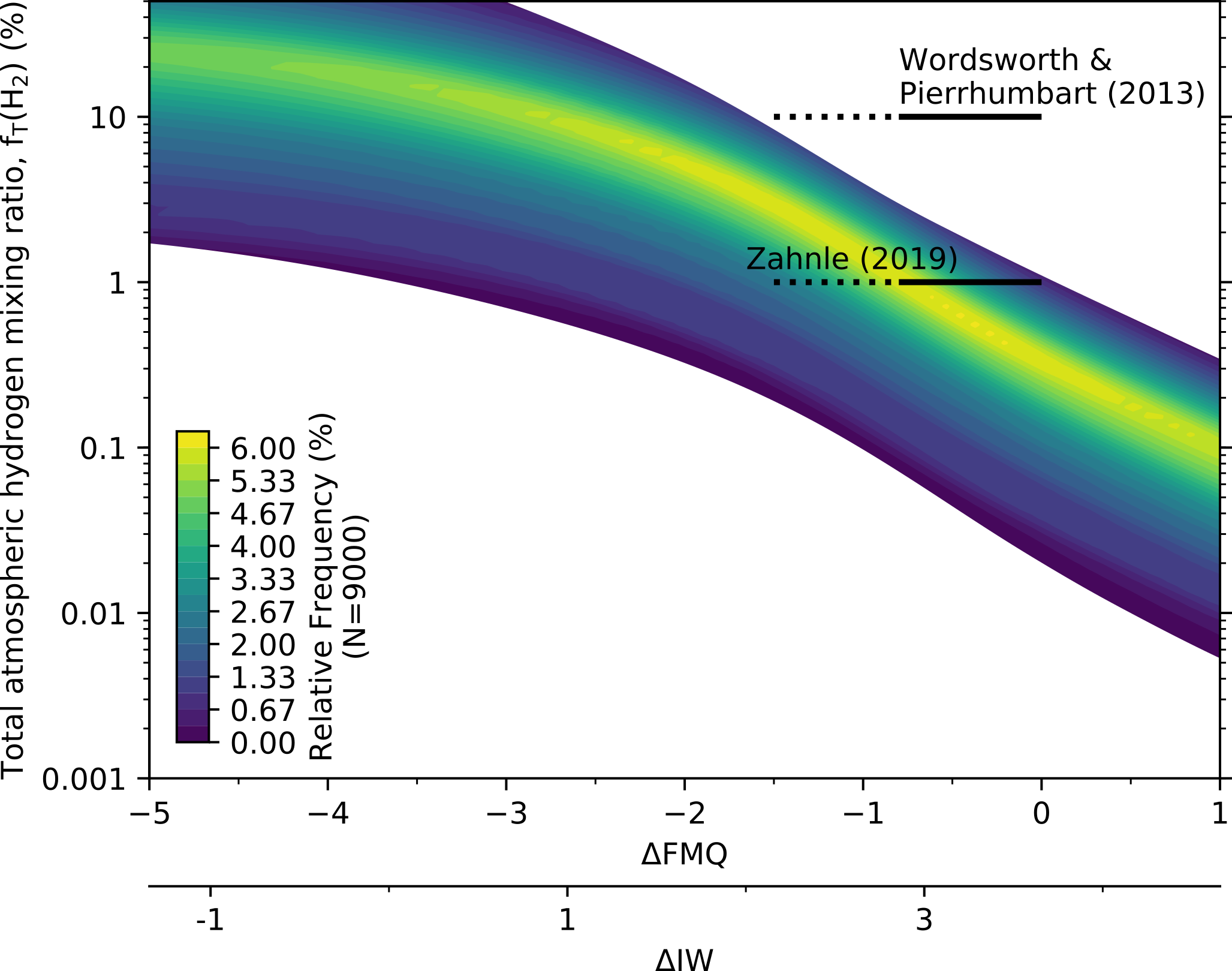}
	  \caption{$f^{\rm v}(\ce{H_2}) + 4f^{\rm v}(\ce{CH4)} + 3f^{\rm v}(\ce{H2S})$} 
	  \label{fig:prob_e_high}
	\end{subfigure}%
	\caption{The relative frequency of hydrogen mixing ratios in the atmosphere of an Earth-like planet, as a function of the magma \fo{}, represented as a 2D density surface. $\alpha$ = 0.2-0.6, $\beta$ = 1 - 10 and surface pressures of 0.1 - 10 bar as in the main text, so (b) is equal to Fig.\,6. Labelled according to the definition of \fth{} used, and hence the photochemical regime.}
	\label{fig:prob_e_photochem}
\end{center}	
\end{figure}

\begin{figure}[H]
	\begin{center}
	\captionsetup[subfigure]{justification=centering}
	\begin{subfigure}[t]{.3\textwidth}
	  \centering
	  \includegraphics[width=\linewidth]{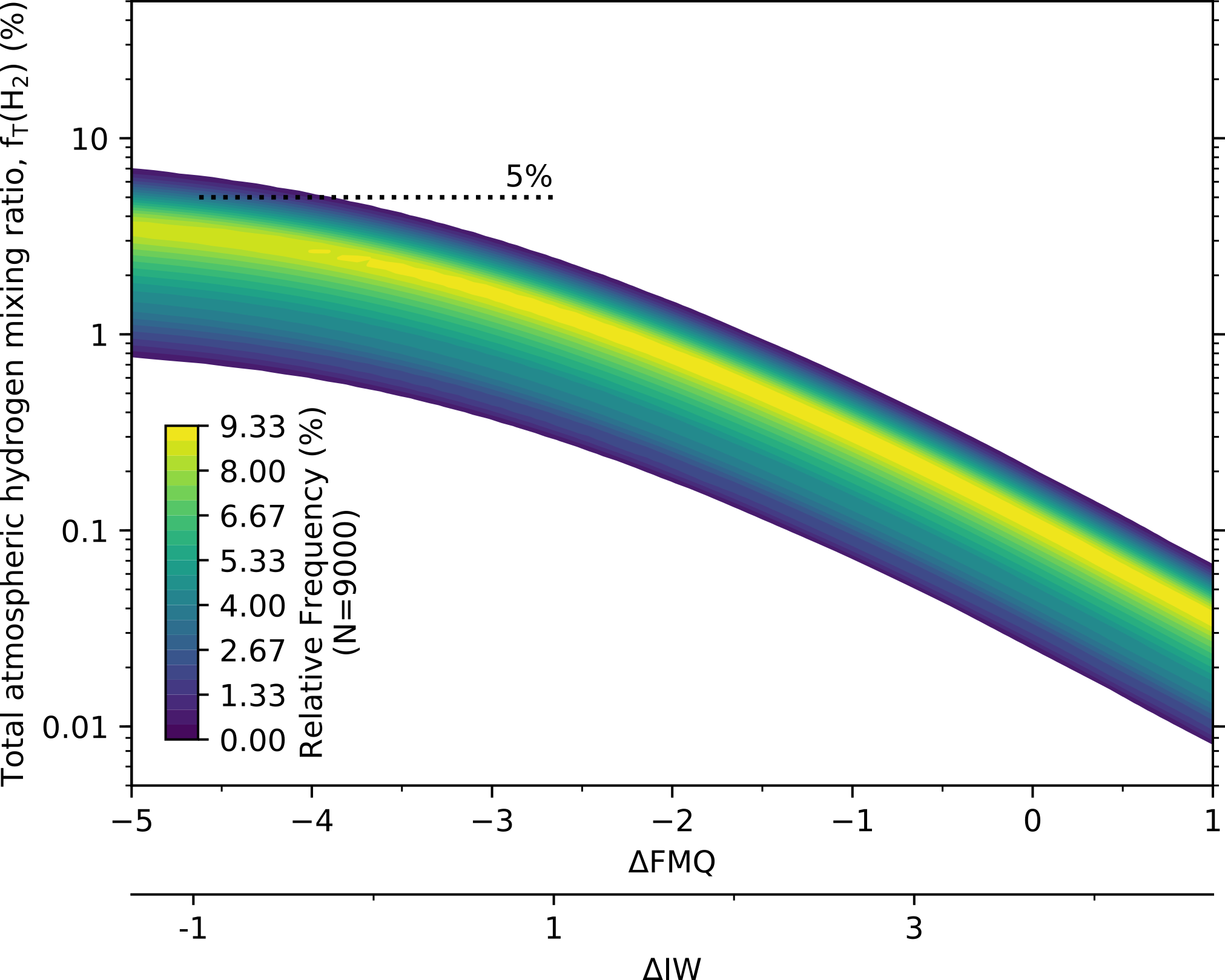}
	  \caption{$f^{\rm v}(\ce{H_2})$} 
	  \label{fig:mars_low}
	\end{subfigure}\hfill%
	\begin{subfigure}[t]{.3\textwidth}
		\centering
		\includegraphics[width=\linewidth]{mars_norm.png}
		\caption{$f^{\rm v}(\ce{H_2}) + 2f^{\rm v}(\ce{CH4)} + f^{\rm v}(\ce{H2S})$} 
		\label{fig:mars_norm}
	  \end{subfigure}\hfill%
	\begin{subfigure}[t]{.3\textwidth}
	  \centering
	  \includegraphics[width=\linewidth]{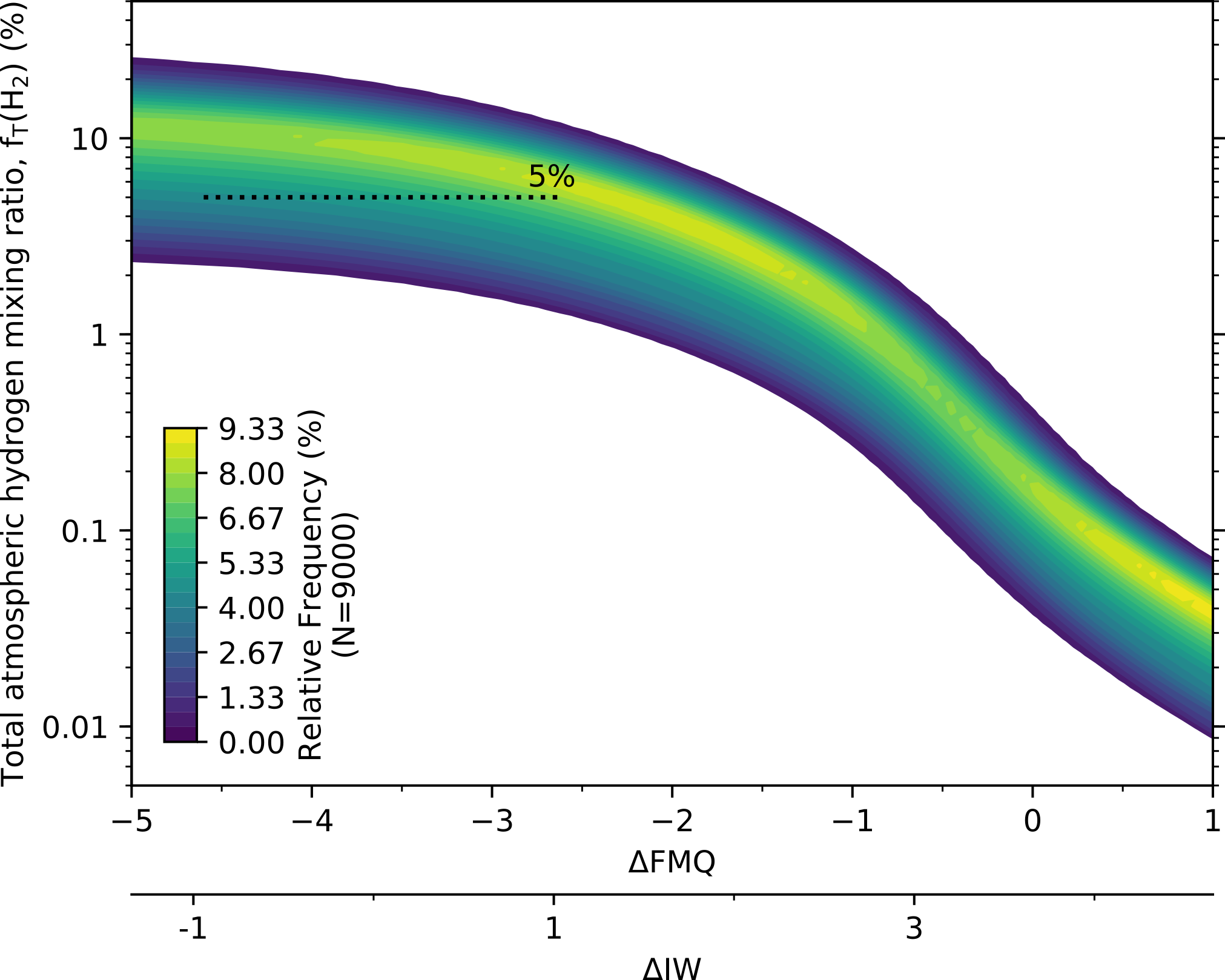}
	  \caption{$f^{\rm v}(\ce{H_2}) + 4f^{\rm v}(\ce{CH4)} + 3f^{\rm v}(\ce{H2S})$} 
	  \label{fig:mars_high}
	\end{subfigure}%
	\caption{The relative frequency of hydrogen mixing ratios in the atmosphere of a Mars-like planet, as a function of the hydrous magma \fo{}, represented as a 2D density surface. $\alpha$ = 0.6\,-\,1, $\beta$ = 0.5 - 2 and surface pressures of 0.5 - 2 bar as in the main text, so (b) is equal to Fig.\,8(a). Labelled according to the definition of \fth{} used, and hence the photochemical regime.}
	\label{fig:prob_m_photochem}
\end{center}	
\end{figure}

To illustrate the effect of different photochemical regimes on our results for the Archean Earth and early Mars, we show in plots \ref{fig:prob_e_low}, \ref{fig:prob_e_norm} and \ref{fig:prob_e_high} for Earth and \ref{fig:mars_low}, \ref{fig:mars_norm} and \ref{fig:mars_high} for Mars the effect of using the three different photochemical regimes.

On Earth, a change in the photochemical regime makes very little difference for oxidised systems above FMQ\,=\,0. The most frequent set of results for achieving 1\,\% \fth{} fall in different areas according to the definition of \ftvh{}, from low \ftvh{} (Fig.\,\ref{fig:prob_e_low}, early \fo{} range) to high \ftvh{} (Fig.\,\ref{fig:prob_e_high}, modern \fo{} range). However, the range in \fth{} at FMQ-1.5 is significantly affected, so where in Fig. \ref{fig:prob_e_low} at FMQ-1.5 the most frequent result is 1-3\,\% \fth{}, in Fig\,\ref{fig:prob_e_high} this range is 2-6\,\%.

For a Mars-like planet with a hydrous melt scenario, a much larger variation is seen over the relevant \fo{} range. At IW+1, the low \ftvh{} regime (Fig\,\ref{fig:mars_low}) has the most frequent results falling into the range 2\,-\,3\,\% \fth{}; at the same \fo{} with a high \ftvh{} regime, this range is 8\,-\,12\,\% \fth{} (Fig\,\ref{fig:mars_high}). At IW-1, these ranges are 3\,-\,5\,\% and 10\,-\,18\,\% \fth{}, for low and high \ftvh{} regimes respectively.

\begin{figure}[H]
	\begin{center}
	\captionsetup[subfigure]{justification=centering}
	\begin{subfigure}[t]{.3\textwidth}
	  \centering
	  \includegraphics[width=\linewidth]{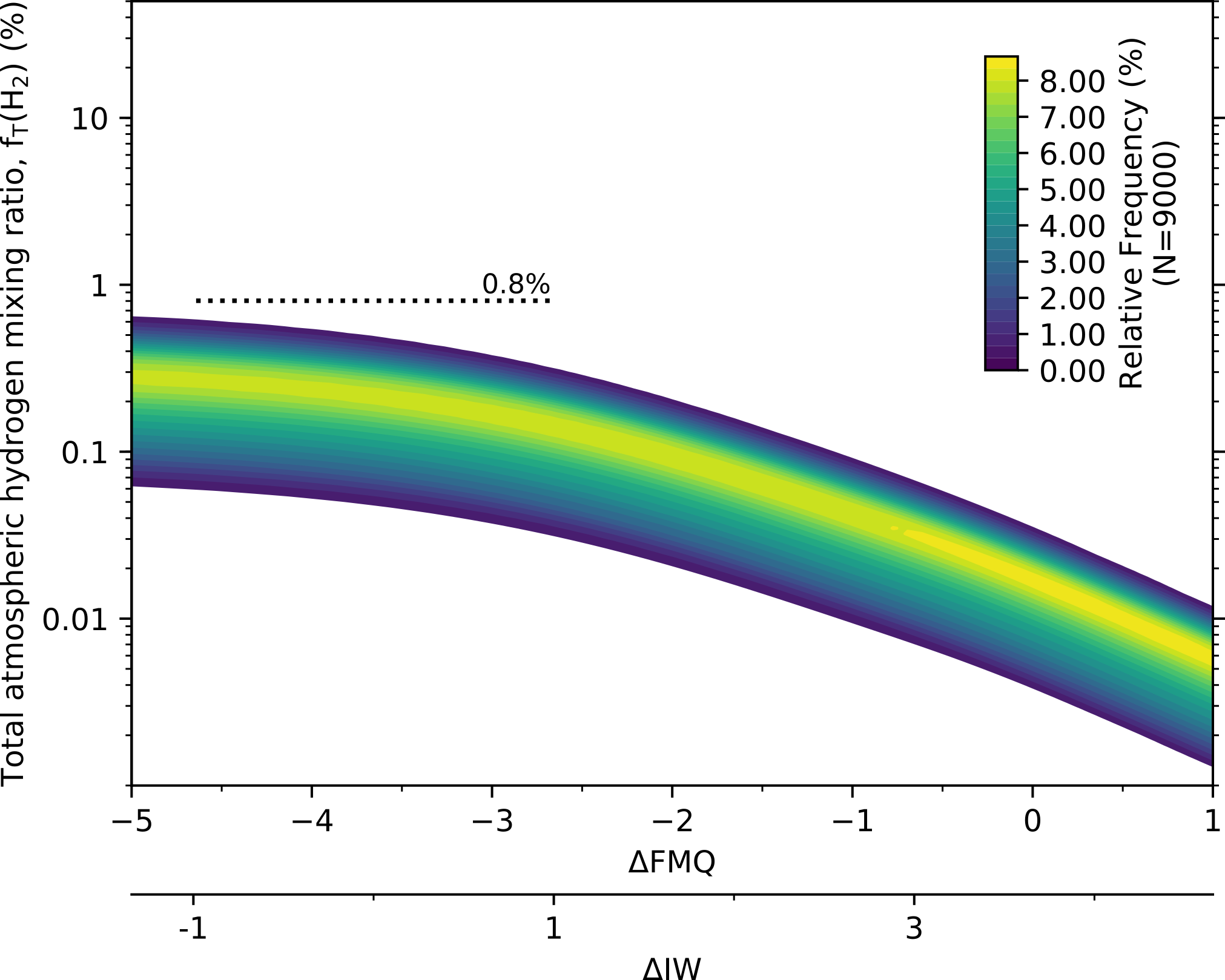}
	  \caption{$f^{\rm v}(\ce{H_2})$} 
	  \label{fig:mars_an_low}
	\end{subfigure}\hfill%
	\begin{subfigure}[t]{.3\textwidth}
		\centering
		\includegraphics[width=\linewidth]{manhydrous_normH.png}
		\caption{$f^{\rm v}(\ce{H_2}) + 2f^{\rm v}(\ce{CH4)} + f^{\rm v}(\ce{H2S})$} 
		\label{fig:mars_an_norm}
	  \end{subfigure}\hfill%
	\begin{subfigure}[t]{.3\textwidth}
	  \centering
	  \includegraphics[width=\linewidth]{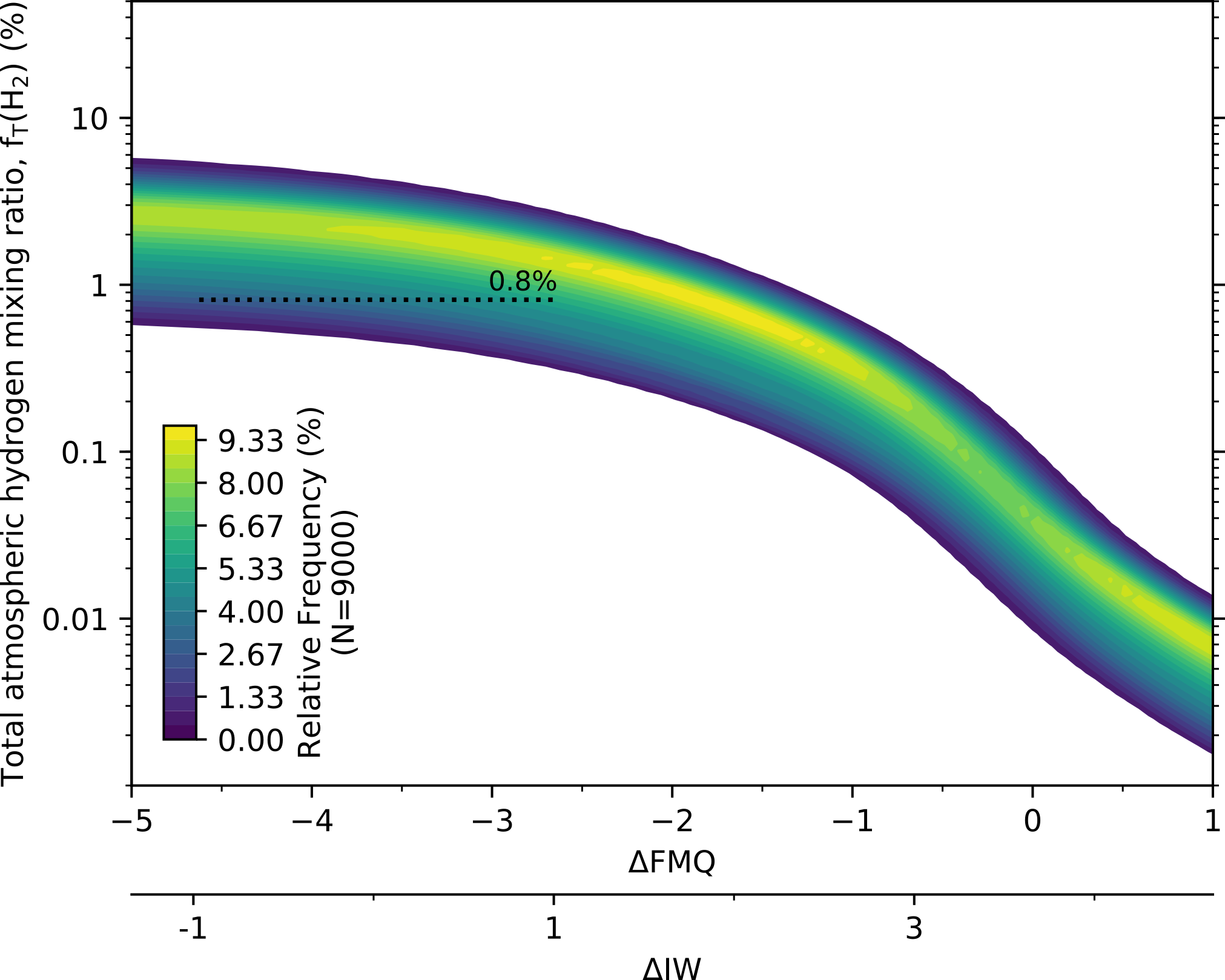}
	  \caption{$f^{\rm v}(\ce{H_2}) + 4f^{\rm v}(\ce{CH4)} + 3f^{\rm v}(\ce{H2S})$} 
	  \label{fig:mars_an_high}
	\end{subfigure}%
	\caption{The relative frequency of hydrogen mixing ratios in the atmosphere of a Mars-like planet, as a function of the anhydrous magma \fo{}, represented as a 2D density surface. $\alpha$ = 0.6\,-\,1, $\beta$ = 0.5 - 2 and surface pressures of 0.5 - 2 bar as in the main text, so (b) is equal to Fig.\,8(c). Labelled according to the definition of \fth{} used, and hence the photochemical regime.}
	\label{fig:prob_m_an_photochem}
\end{center}	
\end{figure}

Under anhydrous conditions, the \fth{} ranges for Mars drop so at IW+1, the low \ftvh{} regime (Fig\,\ref{fig:mars_an_low}) has the most frequent results falling into the range 0.1\,-\,0.2\,\% \fth{}; at the same \fo{} with a high \ftvh{} regime, this range is 1\,-\,2\,\% \fth{} (Fig\,\ref{fig:mars_high}). At IW-1, these ranges are 0.2\,-\,0.35\,\% and 2\,-\,3\,\% \fth{}, for low and high \ftvh{} regimes respectively.